\documentclass[twocolumn,showpacs,preprintnumbers,amsmath,amssymb,superscriptaddress,groupedaddress,prmater]{revtex4}

\usepackage{epsfig}                                                            
\usepackage{graphicx}
\usepackage{dcolumn}
\usepackage{color}
\usepackage{bm}
\usepackage{subfigure}
\newcommand{\be}{\begin{equation}} \newcommand{\ee}{\end{equation}}
\newcommand{\bea}{\begin{eqnarray}} \newcommand{\eea}{\end{eqnarray}}

\begin{document}

\title{\bf Evidence of spin-reorientation transition below 150 K from magnetic force microscopy in ferromagnetic BiFeO$_3$ thin film }

\author {Sudipta Goswami} \email {drsudiptagoswami@gmail.com} \affiliation {School of Materials Science and Nanotechnology, Jadavpur University, Kolkata 700032, India}
\author {Shubhankar Mishra} \affiliation {Department of Physics, Indian Institute of Science, Bangalore 560012, India }
\author {Kishor Kumar Sahoo} \affiliation {Department of Materials Science and Engineering, Indian Institute of Technology, Kanpur 208016, India}
\author {Kumar Brajesh} \affiliation {Department of Materials Science and Engineering, Indian Institute of Technology, Kanpur 208016, India}
\author {Mihir Ranjan Sahoo} \affiliation {Rheinland-Pfalzische Technische Universitat Kaiserslautern-Landau, Erwin-Schrodinger-Stra$\beta$e 52, 67663 Kaiserslautern, Germany}
\author {Subhashree Chatterjee} \affiliation {School of Physical Sciences, Indian Association for the Cultivation of Science, Kolkata 700032, India}
\author {Devajyoti Mukherjee} \affiliation {School of Physical Sciences, Indian Association for the Cultivation of Science, Kolkata 700032, India}
\author {Kalpataru Pradhan} \affiliation {Theory Division, Saha Institute of Nuclear Physics, A CI of HBNI, Kolkata 700064, India} 
\author {Ashish Garg} \affiliation {Department of Materials Science and Engineering, Indian Institute of Technology, Kanpur 208016, India} \affiliation {Department of Sustainable Energy Engineering, Indian Institute of Technology, Kanpur 208016, India}
\author {Chandan Kumar Ghosh} \affiliation {School of Materials Science and Nanotechnology, Jadavpur University, Kolkata 700032, India}
\author {Dipten Bhattacharya} \affiliation {Advanced Materials and Chemical Characterization Division, CSIR-Central Glass and Ceramic Research Institute, Kolkata 700032, India}

\date{\today}

\begin{abstract}
We investigated the magnetic transitions in BiFeO$_3$ at low temperature (5-300 K) and observed nearly 90$^o$ rotation of magnetic domains (imaged by vertical magnetic force microscopy) across 150 K in an epitaxial thin film of thickness $\sim$36 nm. It offers a clear evidence of spin-reorientation transition. It also corroborates the transition observed below $\sim$150 K in the zero-field-cooled and field-cooled magnetization versus temperature data. The field-driven 180$^o$ domain switching at room temperature, on the other hand, signifies presence of ferromagnetism. Since bulk antiferromagnetic BiFeO$_3$ does not exhibit such a transition, this observation in ferromagnetic thin film of BiFeO$_3$ indicates a radical effect because of epitaxial strain. Density functional theory based first-principles calculations too reveal that combined in- and out-of-plane epitaxial strain induces magnetic transition from G- to C-type structure in BiFeO$_3$. 
\end{abstract}
\maketitle

\section{Introduction}
In spite of extensive research on BiFeO$_3$ during the last two decades, consensus is yet to be reached on its different magnetic and crystallographic structures across a wide temperature range 5-1273 K. For example, whether BiFeO$_3$ indeed exhibits a series of magnetic transitions below room temperature - following the transition at $T_N$ $\approx$ 670 K - is still an unsolved issue. Especially, the transition around 150 K created quite a bit of interest \cite{Singh-1,Singh-2,Singh-3,Scott-1,Scott-2,Ramazanoglu}. It was first detected by Singh $\textit{et al}$. [Ref. 2]. The Raman data indicated anomalies in magnon excitations around 150 K \cite{Singh-2,Singh-3,Scott-1}. However, whether the transition is genuinely magnetic or structural could not be resolved beyond reasonable doubt. Later, the transition near 150 K was shown \cite{Scott-3} to be occuring only on the surface of the nanorods of BiFeO$_3$ but not in the single crystal. Therefore, the transition was considered a surface phenomenon. Indeed, early \cite{Fischer} as well as recent \cite{Scott-2,Ramazanoglu} temperature-dependent neutron diffraction studies on a single crystal of BiFeO$_3$ could not detect any magnetic phase transition around 150 K. The magnetic moment in bulk BiFeO$_3$ (space group $R3c$) is contained within the (111) plane and associated with antiferrodistortive rotation of FeO$_6$ octahedra around the polarization axis [111] \cite{Spaldin}. Stabilization of the G-type structure requires canting of the moment by nearly 1$^o$. It also assumes spiral modulation over a length scale $\sim$62 nm. Although some work have been done \cite{Huang,Bertinshaw,Gervits} to examine the magnetic structure, especially, the spin cycloid - its length, propagation vector, role of strain and doping etc, no detailed work has, so far, been done to investigate the size dependence of the magnetic structure across a wide size range and, more importantly, the possible magnetic transitions and stabilization of different magnetic phases across the entire temperature range.  

\begin{figure}[ht!]
\centering
{\includegraphics[scale=0.30]{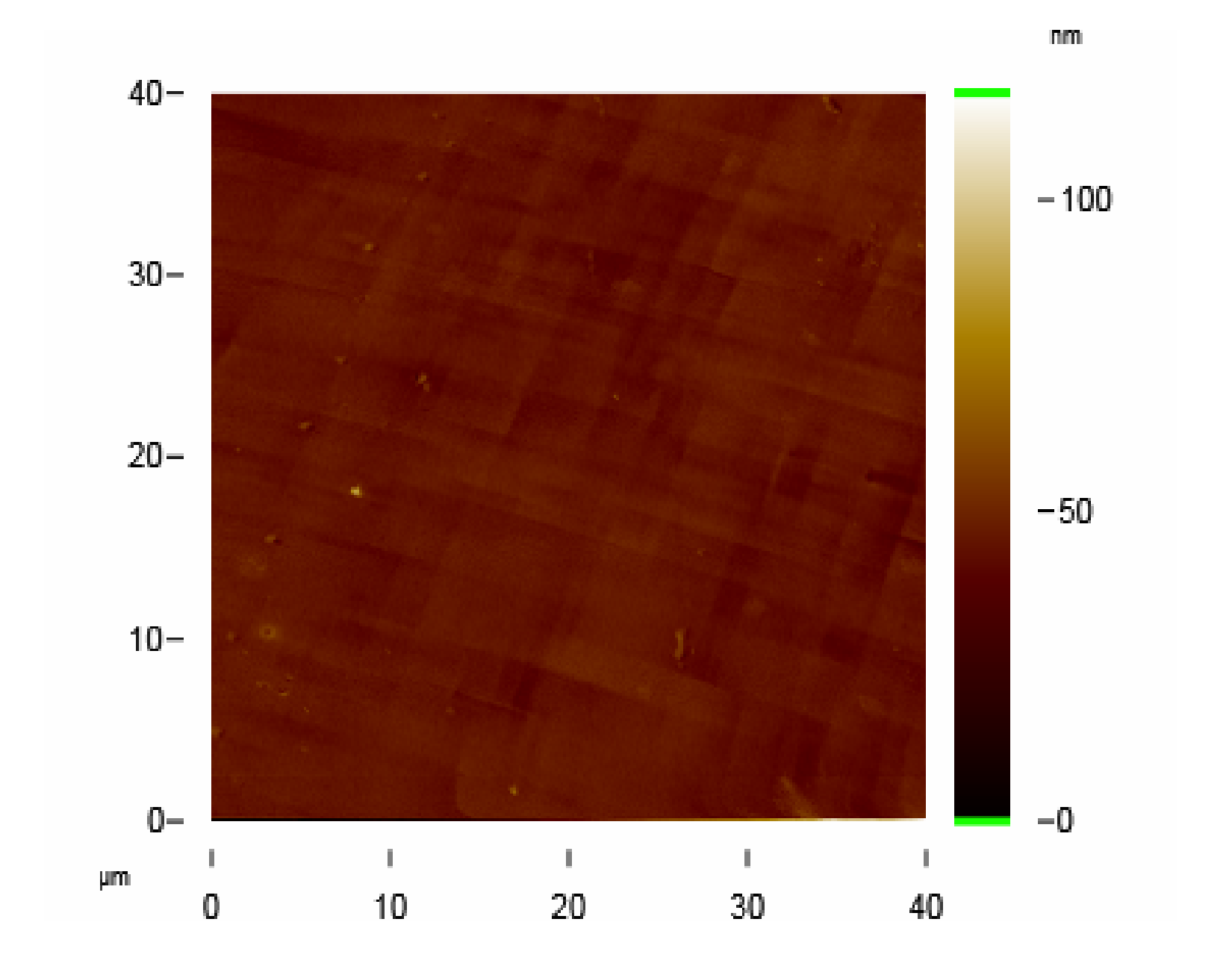}}
\caption{The room temperature surface topography image of the BiFeO$_3$ film.}
\end{figure}

\begin{figure*}[ht!]
\centering
{\includegraphics[scale=0.45]{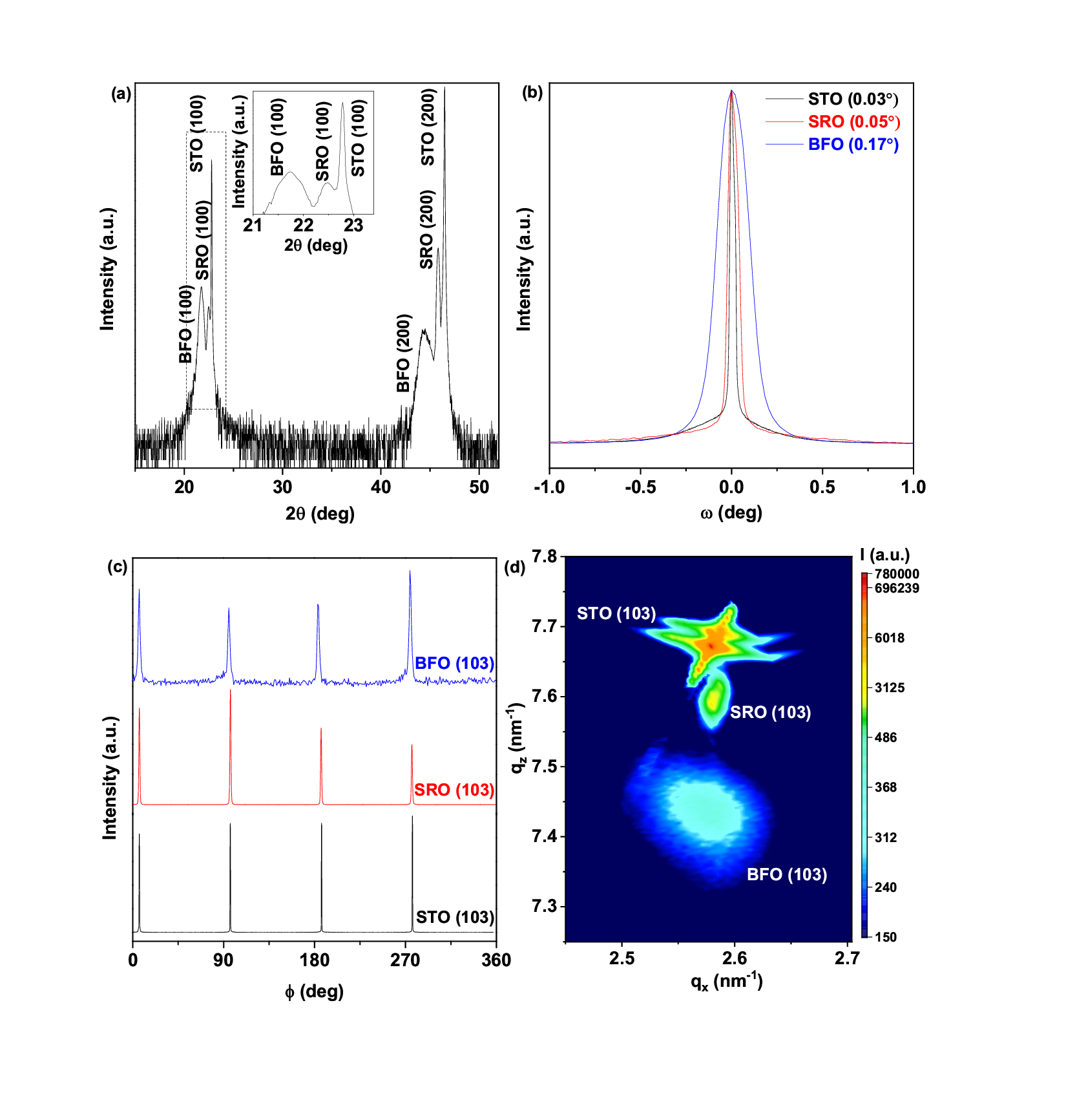}}
\caption{The room temperature (a) XRD $\theta-2\theta$ pattern of BiFeO$_3$ (BFO)/SrRuO$_3$ (SRO)/SrTiO$_3$ (STO) thin film heterostructure; Inset to (a) shows the zoomed-in view aroundthe (100) peak positions of BFO, SRO and STO layers for visual clarity; (b) rocking curves ($\omega$ scans) about the (100) planes for STO (0.03$^o$), SRO (0.05$^o$) and BFO (0.17$^o$) with the full-width half maxima (FWHM) values given in parenthesis;(c) azimuthal ($\phi$) scans about (103) planes of STO, SRO and BFO layers; (d) reciprocal space map captured in the vicinity of asymmetric (103) reflection of STO for the BFO/SRO/STO heterostructure.}
\end{figure*}

Given this backdrop, we examined the issue of magnetic transition at 150 K afresh on a high quality epitaxial thin film of BiFeO$_3$ (thickness $\approx$ 36 nm) grown on (100)SrTiO$_3$ substrate by pulsed laser deposition technique. We observed significantly large anomalous features in the zero-field-cooled and field-cooled (ZFC and FC) magnetization ($M$) versus temperature ($T$) data around 150 K. We also employed magnetic force microscopy (MFM) and obtained clear evidence of domain reorientation by nearly 90$^o$ across the transition. In addition, we demonstrate using density functional theory (DFT) based first-principles calculations that magnetic transition from G- to C-type structure is indeed possible in BiFeO$_3$ as a result of biaxial strain.

\section{Experimental and Computational Details}
The epitaxial thin film of BiFeO$_3$ was prepared by pulsed laser deposition technique on (100)SrTiO$_3$ (STO) substrate. A thin (100)SrRuO$_3$ buffer layer was grown onto the (100)SrTiO$_3$ substrate prior to the growth of the BiFeO$_3$ film. The details of the film growth are available in an earlier paper \cite{Garg}. The surface features of the film was characterized by atomic force microscopy (AFM). The crystallographic orientation of the heterostructure was examined using x-ray diffraction experiments carried out at the Rigaku Smart Lab 9 kW XG diffractometer provided with a five-axis goniometer sample stage. The collimated Cu k$\alpha$ beam (wavelength = 1.5406 \AA) was used. The thickness of the film was measured by a surface profilometer. The ZFC and FC magnetization ($M$) versus temperature ($T$) data were recorded by the SQUID magnetometer (Quantum Design). The room temperature magnetic hysteresis loops were measured by a Vibrating Sample Magnetometer (LakeShore Model 7407). The temperature and the magnetic field dependent magnetic force microscopy (MFM) was carried out by the LT AFM/MFM system of Nanomagnetics Instruments Ltd, Ankara, Turkey. 

\begin{figure*}[ht!]
\begin{center}
   \subfigure[]{\includegraphics[scale=0.20]{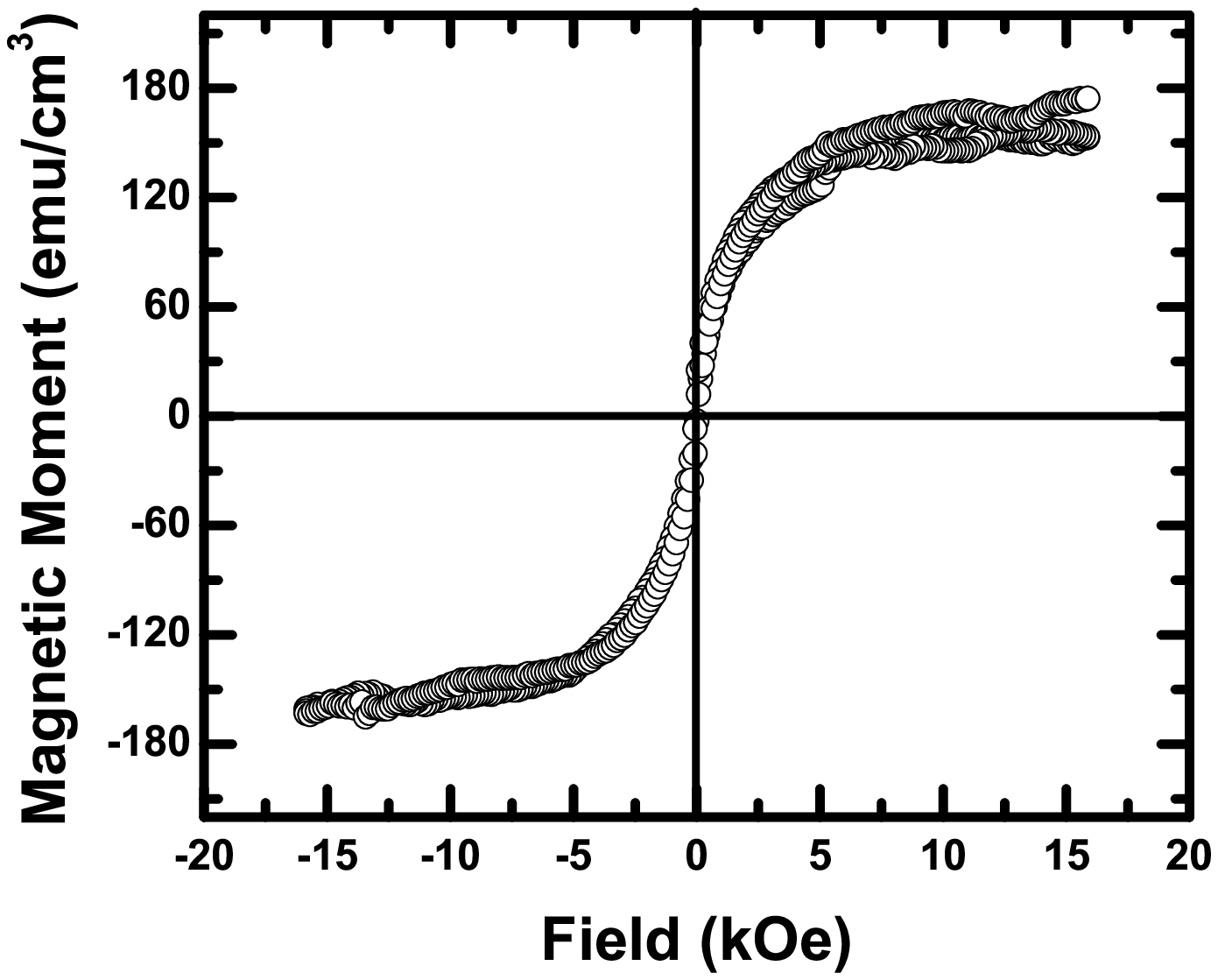}} 
   \subfigure[]{\includegraphics[scale=0.20]{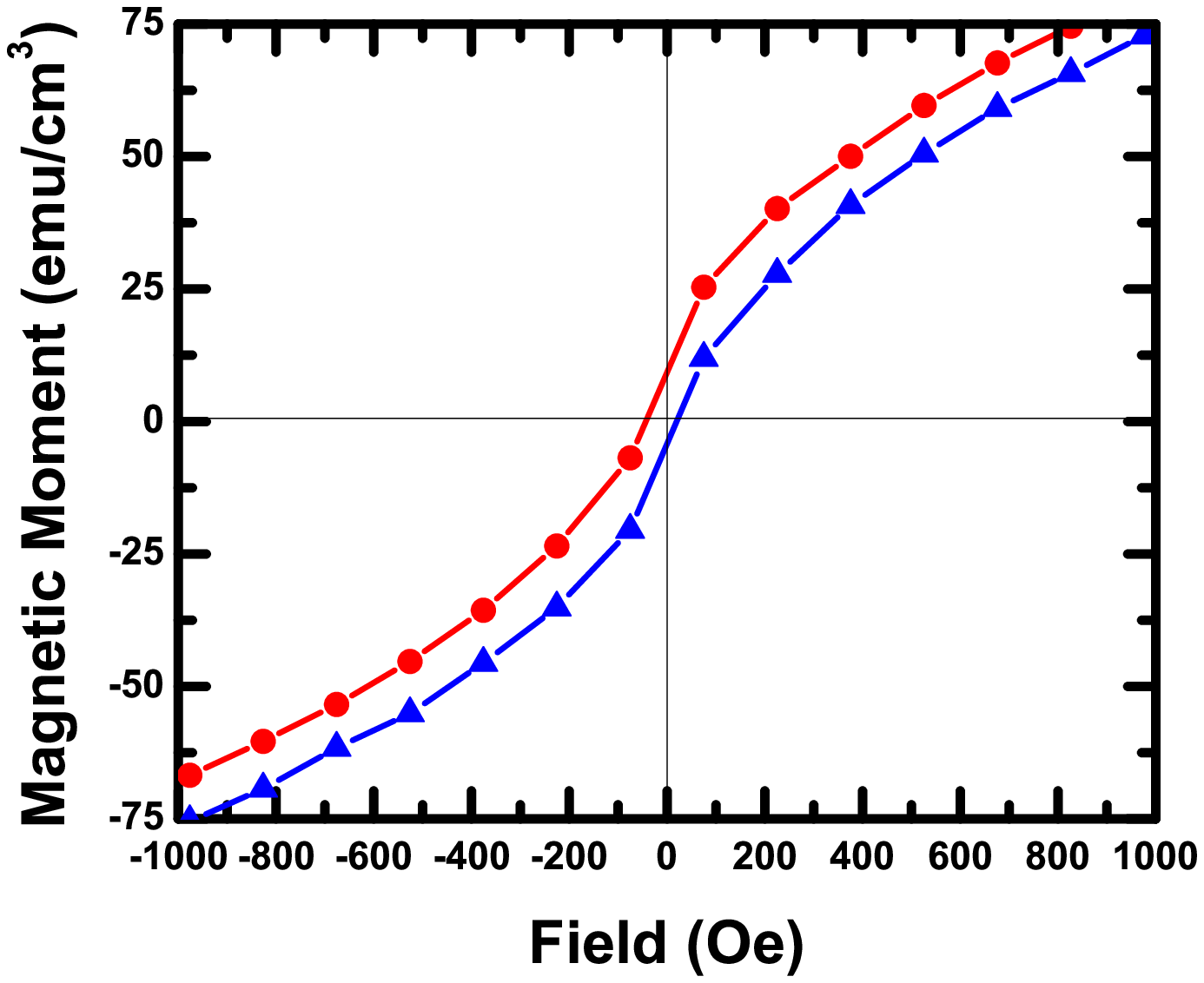}}  
 	\subfigure[]{\includegraphics[scale=0.20]{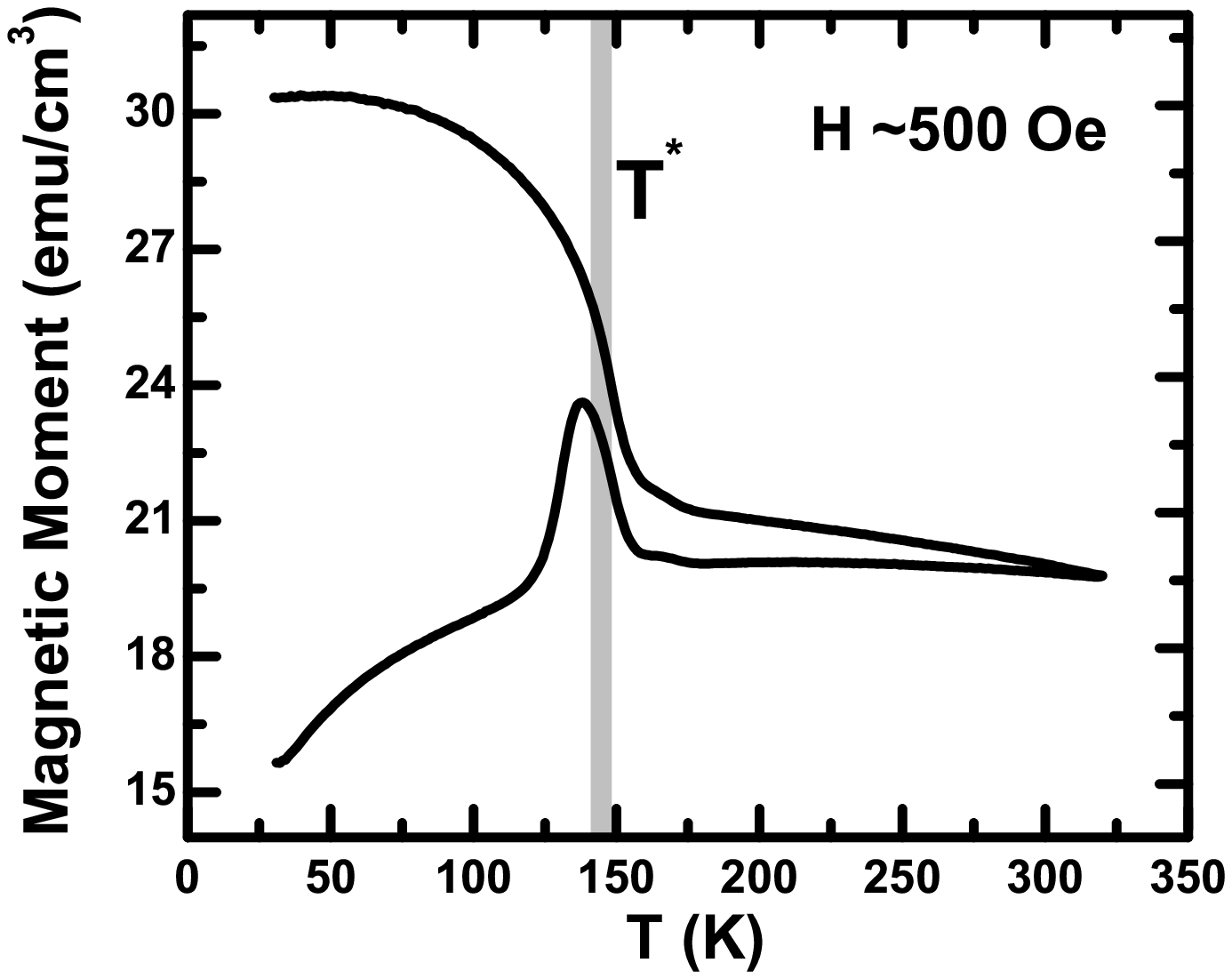}}  
   \end{center}
\caption{The (a) room temperature magnetic hysteresis loop and (b) the portion near the origin of the hysteresis loop has been blown up to show finite coercivity; (c) the ZFC and FC magnetization versus temperature data which show magnetic transition around 150 K.}
\end{figure*}

We carried out band-structure calculations to understand the underlying magnetic structure of BiFeO$_3$. The first principles density functional calculations are performed using Vienna ab initio Simulation Package (VASP) code \cite{Georg,Furthmuller,Joubert} for optimizing the crystallographic structure and analyzing the electronic structure. To describe the electron-ion interactions, the projector-augmented wave (PAW) potential is taken into account. The Perdew-Burke-Ernzerhof (PBE)\cite{Perdew} version of generalized gradient approximation (GGA) is considered as the exchange-correlation functional. The Bi 6s$^2$ 6p$^3$, Fe 3d$^6$ 4s$^2$, and O 2s$^2$ 2p$^4$ electrons are used for the calculations. A kinetic energy cut-off of 500 eV is chosen for the expansion of the plane-wave basis set. The k-point grids of 5 $\times$ 5 $\times$ 5 and 7 $\times$ 7 $\times$ 7 are chosen to sample the first Brillouin zone for spin-polarized self-consistent and density of states (DOS) calculations, respectively. The convergence criterion for the electronic self-consistency is set at 10$^{-5}$ eV. All the atoms in the supercell are relaxed until the Hellman-Feynman forces on each atom are less than 0.001 eV/\AA. Considering the strong Coulomb interaction ($U$) of localized d-states of Fe atoms, GGA+U approach, described by Dudarev $\textit{et al}$. \cite{Dudarev}, is applied to include the correlation effect in transition metal oxides. In this work, the effective Hubbard parameter ($U_{eff}$ = $U-V$) is set at 4.6 eV for Fe atoms only \cite{Walden}.

\section{Results and Discussion}
Figure 1 shows the surface topography image of the film surface. The surface roughness is found to be varying within $\sim$1.0-2.0 nm. Additional images of the surface topography as well as the corresponding data on the surface roughness ($R_a$ values) are available in the supplemental materials document \cite{supplementary}. The single crystalline nature of the BiFeO$_3$ (BFO) and SrRuO$_3$ (SRO) phases in the BFO/SRO heterostructure grown on STO (100) substrate is evidenced from the XRD $\theta-2\theta$ pattern shown in Fig. 2(a). Only strong ($l$00)($l$ = 1, 2, and 3) diffraction peaks of the pseudocubic BFO ($a_0$ = 3.96 \AA), SRO ($a$ = 3.87 \AA, JCPDS 01-071-5344), and STO (cubic, $a$ = 3.905 \AA) phases could be observed. It confirms the cube-on-cube epitaxial growth with no trace of impurity within the resolution limit of the XRD \cite{Botea,Achenbach,Yang}. Since ensuring phase-purity of the nanoscale or epitaxial thin films of BiFeO$_3$ is extremely important, as often presence of minor amoung of ferromagnetic secondary phases (e.g., Fe$_2$O$_3$) could change the magnetic and other physical properties significantly \cite{Bea}, we have carried out the XRD scan across a wider 2$\theta$ range (10$^o$-100$^o$) in order to examine the presence of impurity phases in these thin films, if any. The result is shown in the supplemental materials document \cite{supplementary}. Peaks corresponding to any additional impurity phase could not be observed. Therefore, the film is free from any secondary impurity phase. The out-of-plane lattice parameters ($a_{\perp}$) of the BFO ($a_{\perp}$ = 4.07$\pm$0.02 \AA) and the SRO ($a_{\perp}$ = 3.95$\pm$0.01 \AA) layers were calculated from the $\theta-2\theta$ pattern. Due to the close proximity of the BFO (100), SRO (100) and STO(100) peaks, a close-up view of the BFO/SRO heterostructure is shown in the inset of Fig. 2(a). The rocking curves ($\omega$ scan) [Fig. 2(b)], recorded about the (100) planes of the BFO and SRO layers and STO substrate, exhibit full width at half maxima values to be $\sim$0.17$^o$, 0.05$^o$, and 0.03$^o$, respectively. It confirms high degree of in-plane orientation of the SRO and BFO layers in the BFO/SRO heterostructure. The XRD $\phi$ or azimuthal scans [Fig. 2(c)], performed about the STO (103), SRO (103), and BFO (103) planes, exhibit sharp peaks at regular intervals of 90$^o$ which signifies four-fold symmetry and cube-on-cube epitaxial growth of each layers. The in- and out-of-plane unit cell parameters ($a_{\parallel}$, $a_{\perp}$) were determined from the high-resolution XRD reciprocal space maps (RSM) obtained by using asymmetric reflections. Figure 2(d) shows a representative RSM performed about the STO (103) peak for the BFO/SRO heterostructure \cite{Qi}. Single peaks corresponding to SRO (103) and BFO (103) reflections are observed near the STO substrate peak which signifies the epitaxial growth of both the layers \cite{Botea}. The calculated lattice parameters of BFO ($a_{\parallel}$ = 3.91$\pm$0.010 \AA, $a_{\perp}$ = 4.06$\pm$0.02 \AA) and SRO ($a_{\parallel}$ = 3.91$\pm$0.01 \AA, $a_{\perp}$ = 3.95$\pm$0.01 \AA) layers indicate an in-plane compressive strain, $\epsilon_{\parallel}$ = ($a_{\parallel}$-$a_0$)/$a_0$ $\approx$ -1.26\%, and out-of-plane tensile strain, $\epsilon_{\perp}$ = ($a_{\perp}$-$a_0$)/$a_0$ $\approx$ 2.5\%, in the BFO layer. This is understandable since the BFO layer undergoes an in-plane compression in order to match the slightly smaller lattice parameter of the underlying SRO layer (lattice mismatch of 0.25\%). The $a_{\perp}$ values obtained from the RSM match those obtained from the XRD $\theta-2\theta$ scans [Fig. 2(a)]. The epitaxial strain developed in the BFO layer gives rise to a large tetragonal distortion of the BFO unit cell ($a_{\perp}$/$a_{\parallel}$-1 $\approx$ 3.8\%) which, in turn, could drastically affect its magnetic properties.

\begin{figure*}[ht!]
\begin{center}
   \subfigure[]{\includegraphics[scale=0.15]{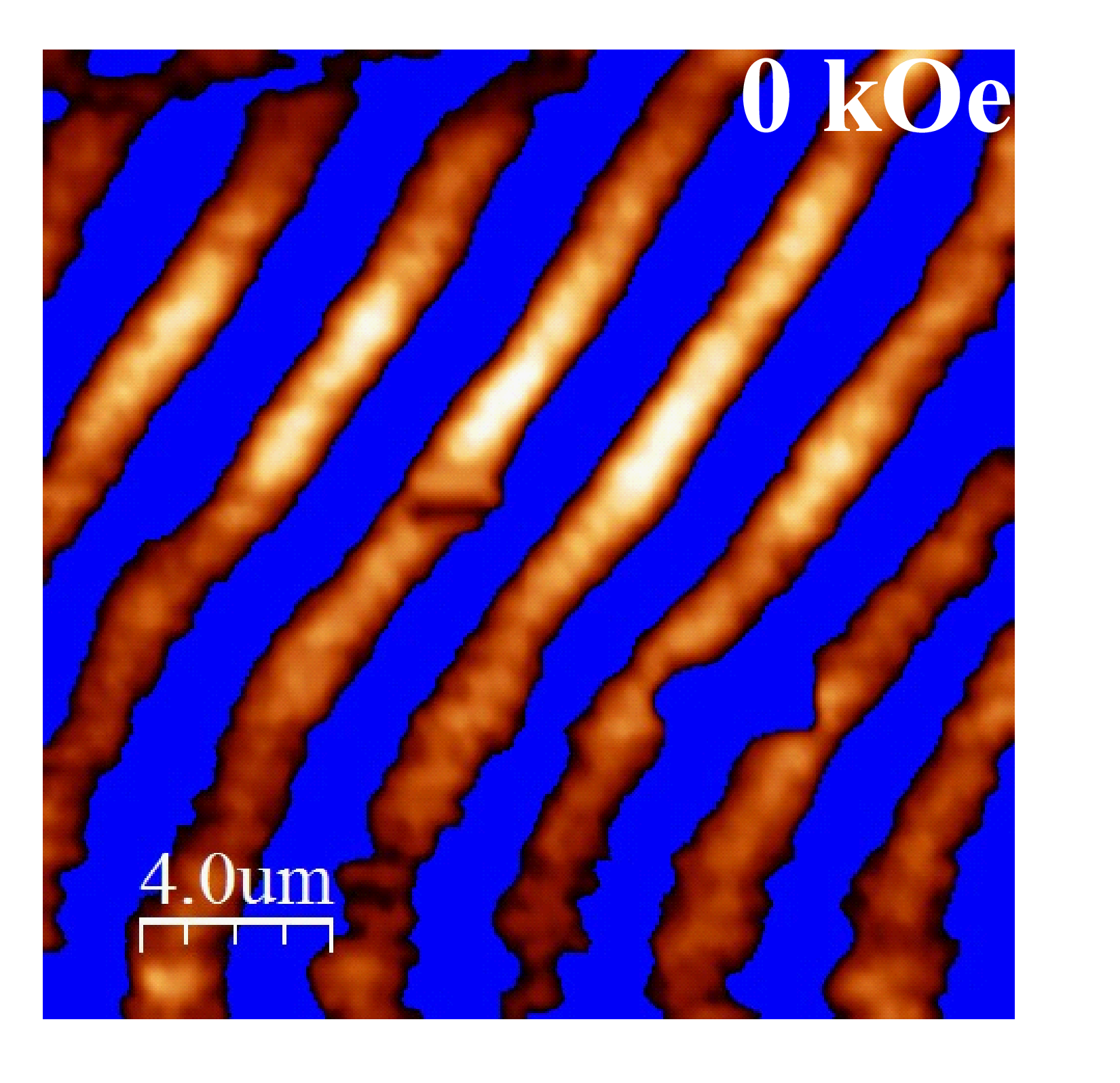}} 
   \subfigure[]{\includegraphics[scale=0.15]{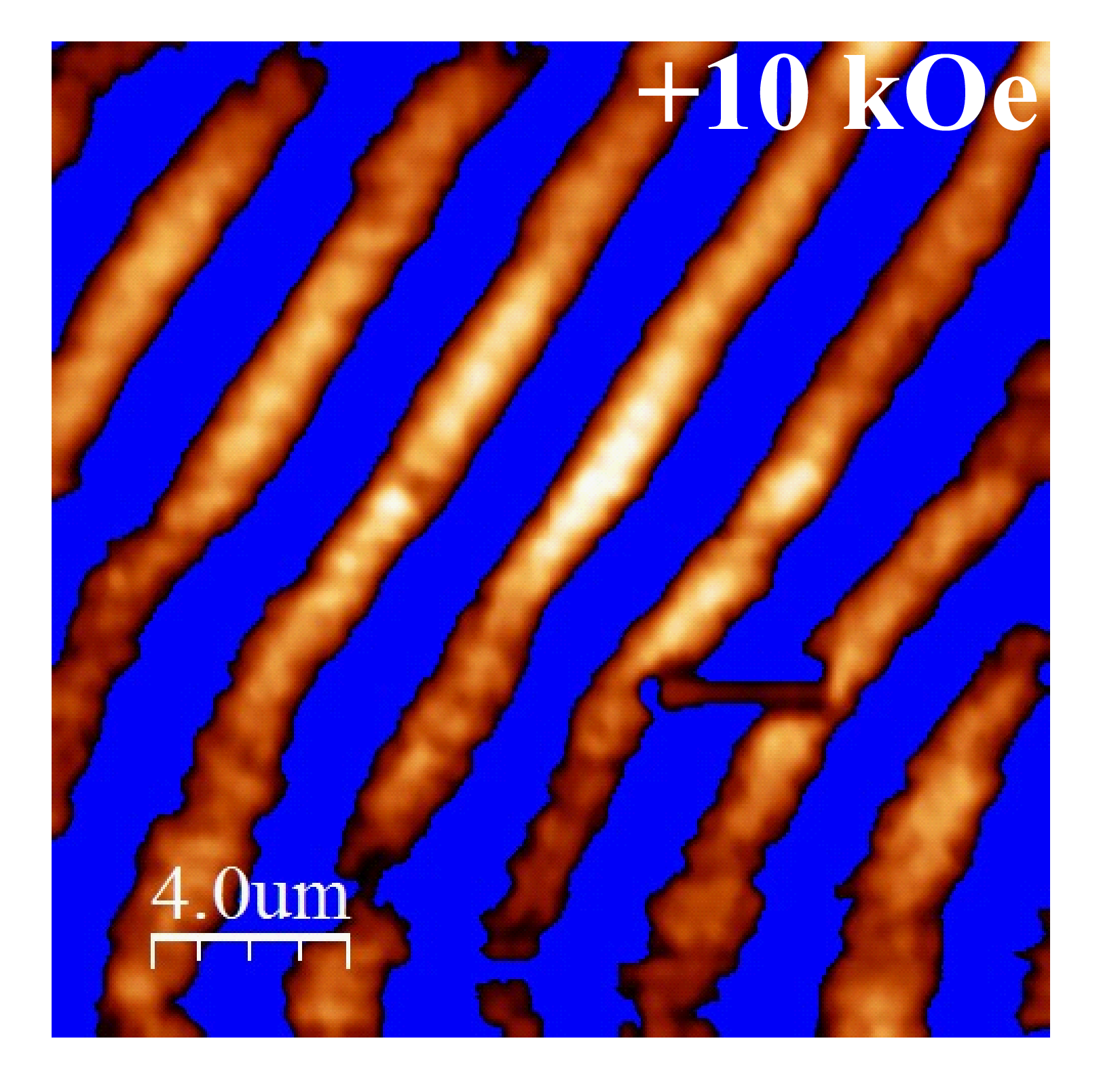}}  
   \subfigure[]{\includegraphics[scale=0.15]{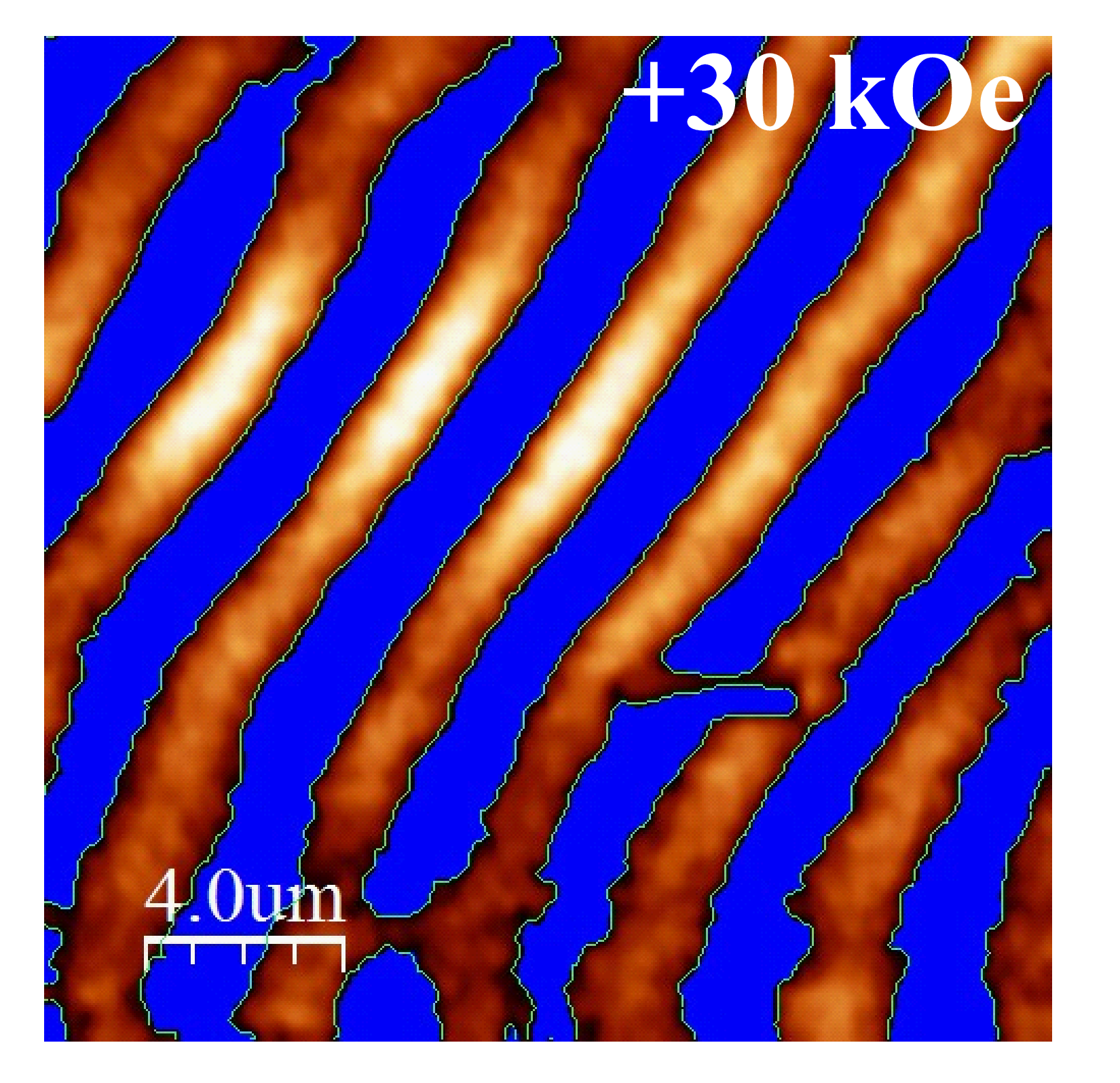}} 
   \subfigure[]{\includegraphics[scale=0.15]{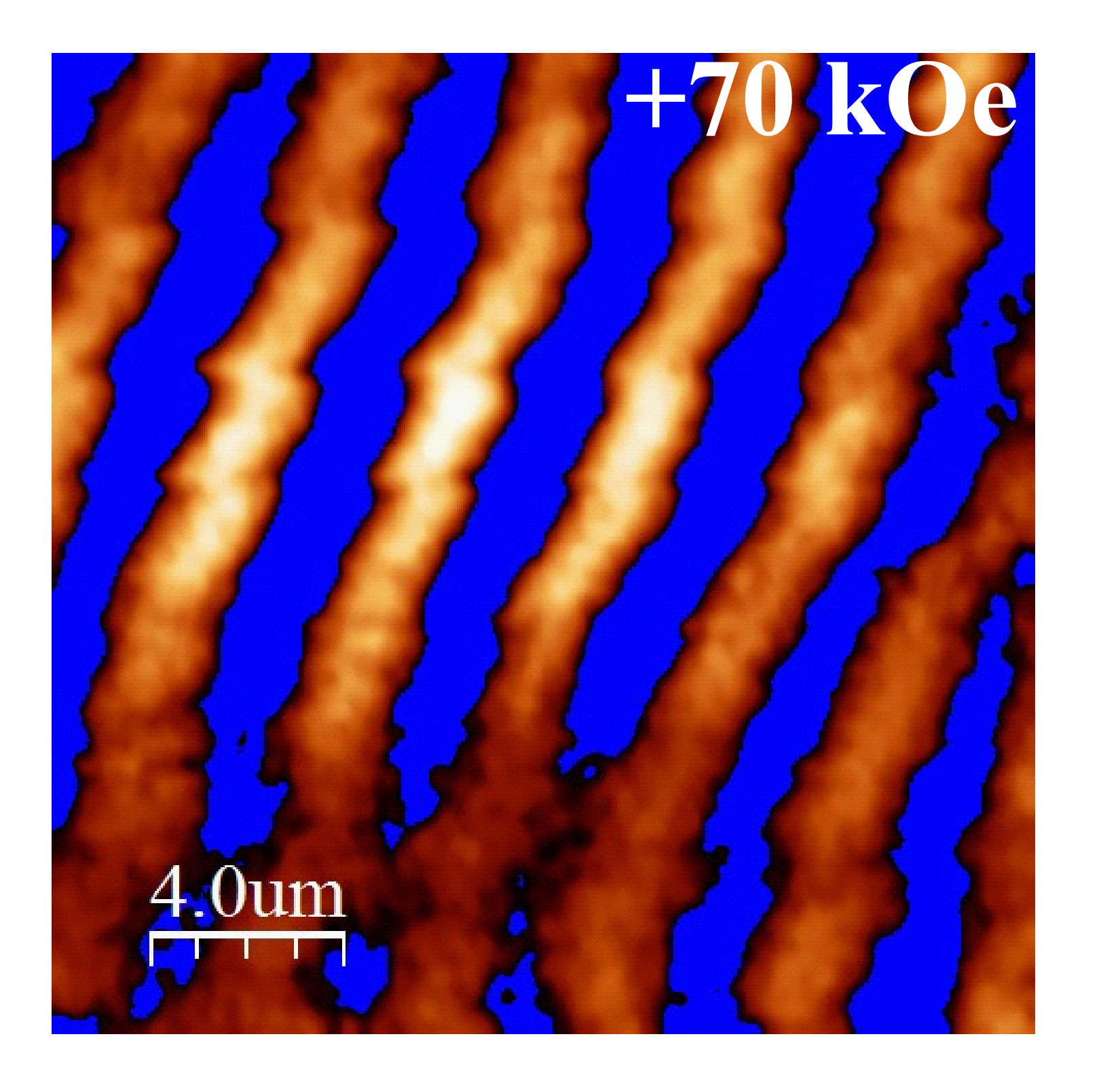}} 
   \subfigure[]{\includegraphics[scale=0.15]{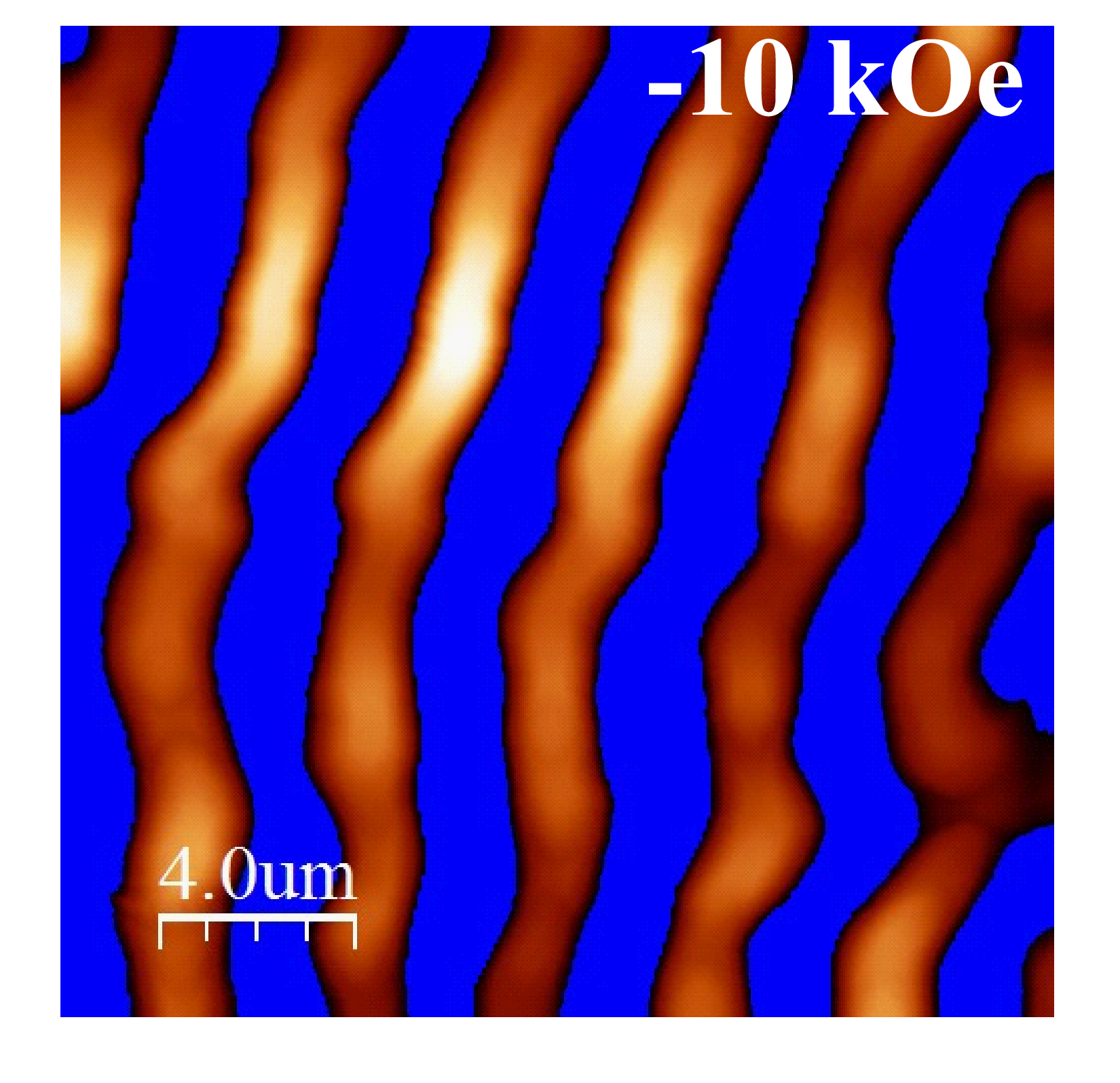}} 
   \subfigure[]{\includegraphics[scale=0.15]{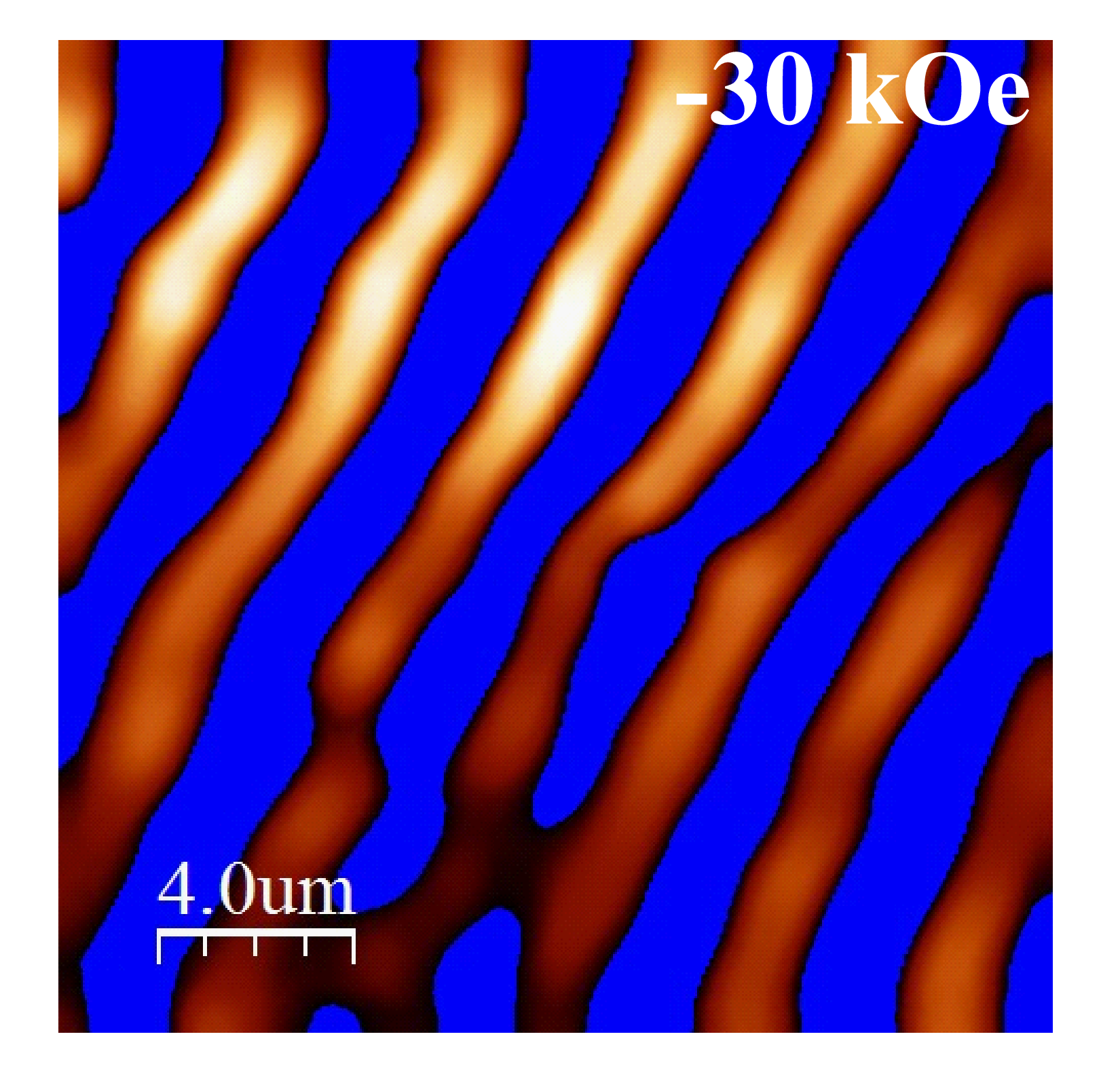}}  
   \subfigure[]{\includegraphics[scale=0.15]{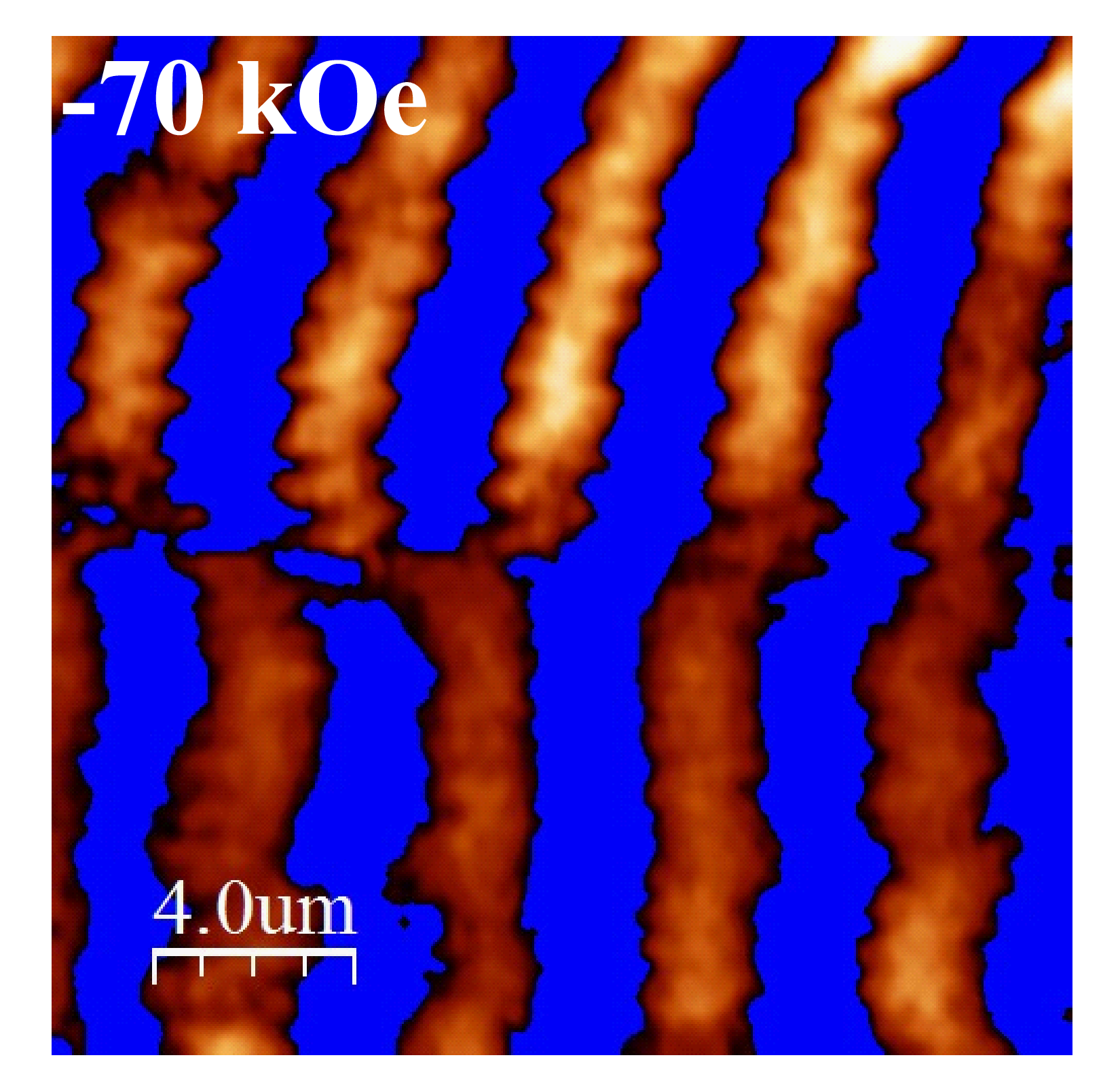}} 
   \subfigure[]{\includegraphics[scale=0.15]{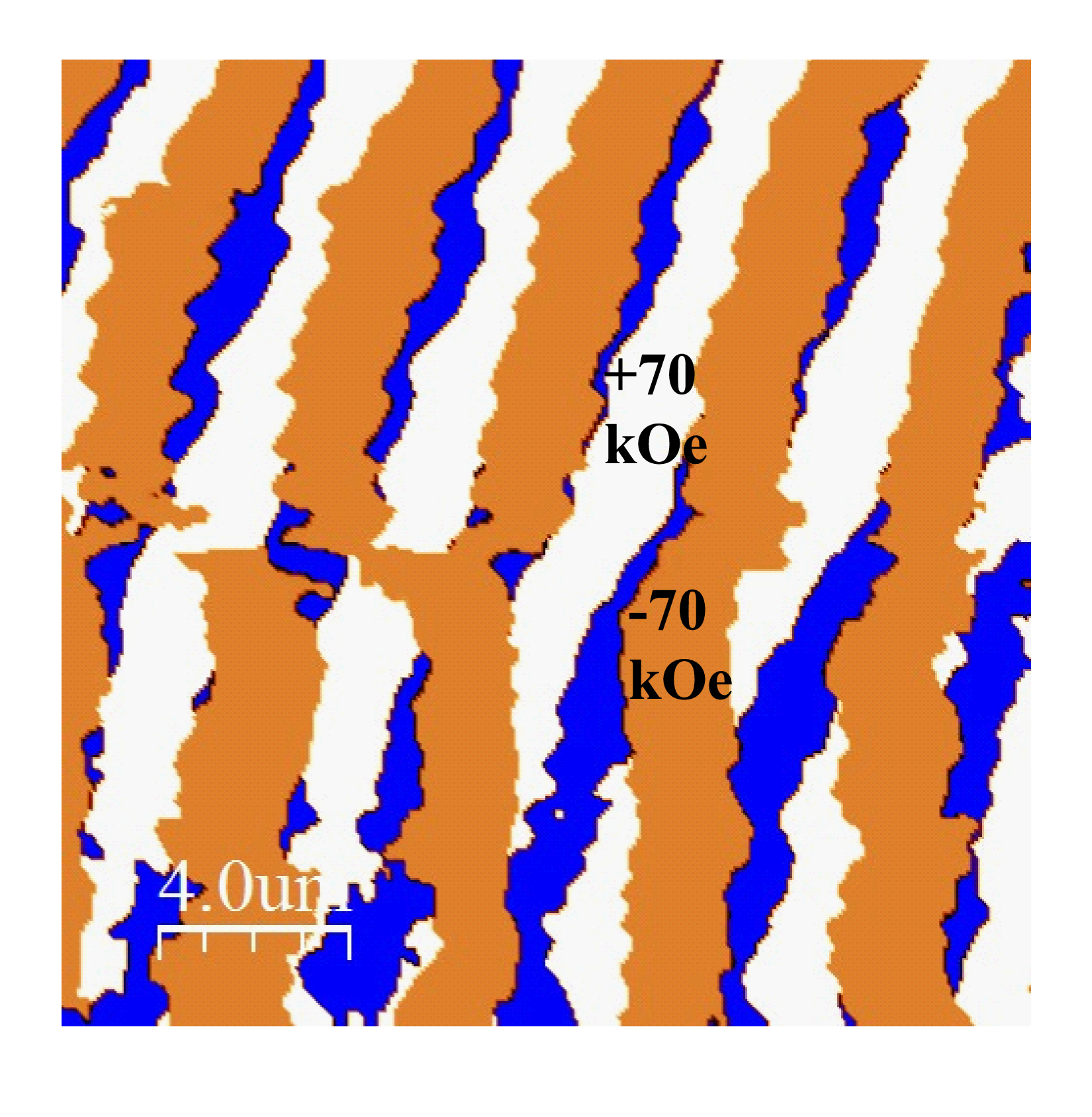}} 
   \subfigure[]{\includegraphics[scale=0.20]{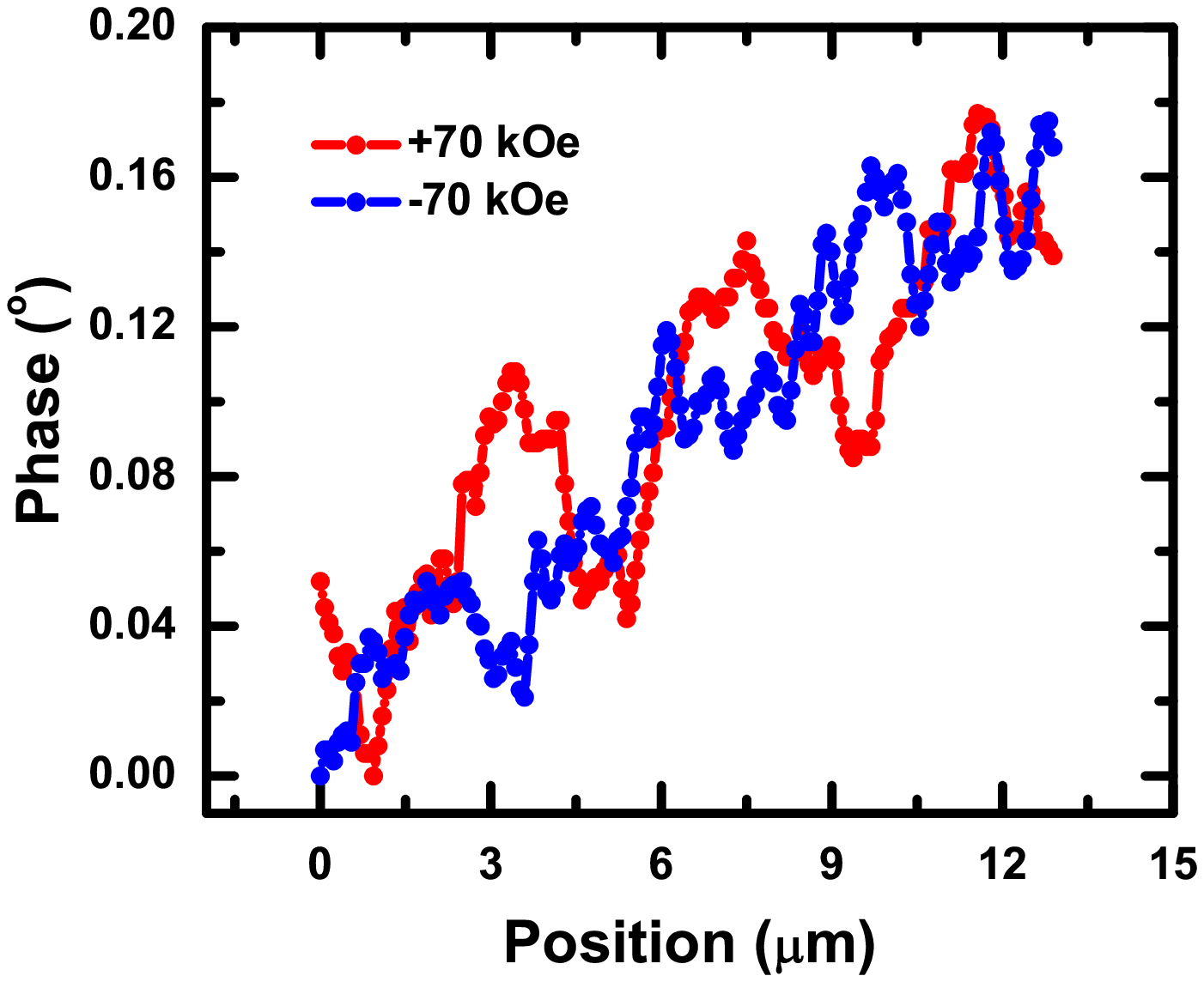}}  
   \subfigure[]{\includegraphics[scale=0.20]{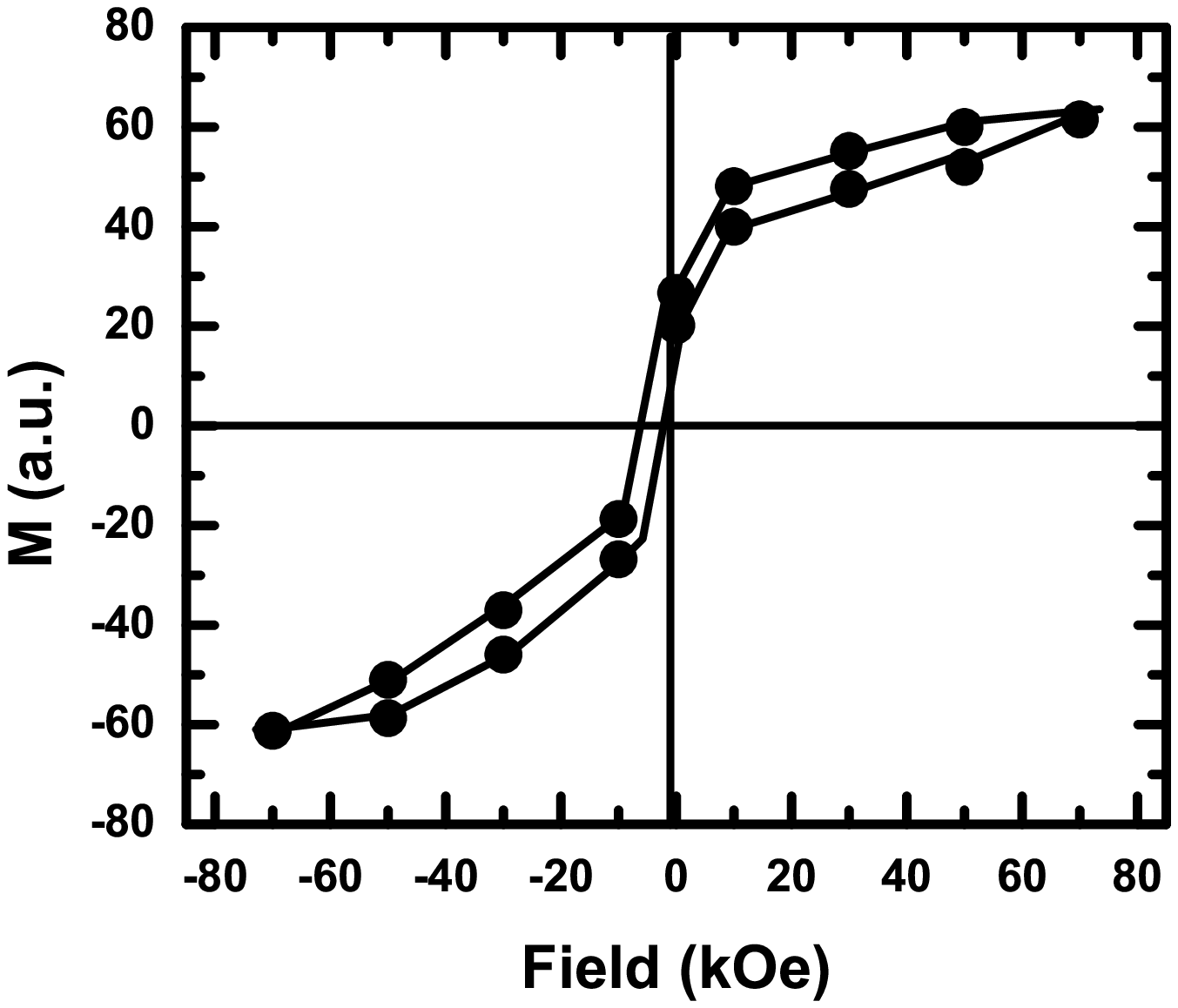}} 
   \end{center}
\caption{(a-g)The phase-contrast room-temperature MFM images under different magnetic field during tracing the forward and reverse branches of the magnetic hysteresis loop; (h) superimposed MFM images captured under +70 and -70 kOe field. (i) line scan data under +70 and -70 kOe field; (j) magnetic hysteresis loop constructed by using the data of switched volume of magnetic domains; all these images have been recorded by scanning the same area of the film surface using vertical MFM (magnetic field is oriented perpendicular to the film surface); the images have been processed by WSxM software; the blue color represents the background; the vertical scale could be noted from the plot of line scan data shown in supplemental materials document \cite{supplementary}.} 
\end{figure*}

We carried out detailed magnetometry on this film. Room temperature magnetic hysteresis loop was measured with the magnetic field applied parallel to the film surface. The contribution of the background - substrates, sample holder, fixture etc. - has been eliminated appropriately. The loop [Fig. 3(a)] signifies presence of soft ferromagnetism; the portion of the loop near the origin has been blown up in Fig. 3(b). The coercivity turns out to be small yet finite ($\sim$50 Oe). The lateral dimensions of the films were 5 mm $\times$ 5 mm. The minimum magnetic moment of the sample was, therefore, of the order of $\sim$10$^{-5}$ emu which is nearly two orders of magnitude higher than the detection limit of the 7407 model vibrating sample magnetometer of LakeShore. Slight mismatch between the initial and final branches of the loop at the positive saturation magnetization end could stem from small demagnetization of the sample prior to the measurement which resulted in slightly lower magnetic induction in the sample at the start of the cycle than that attained at the end. The saturation magnetization $M_S$ is large ($\sim$180 emu/cm$^3$). However, it conforms well to the earlier experimental results \cite{Wang,Ryu,Zhang,Ramazanov,Prokhorov,Mocherla,Reddy,Kartavseva,Park,Mazumder} on enhanced ferromagnetism and large $M_S$ (varying within 50-200 emu/cm$^3$) for epitaxial thin films and nanoscale BiFeO$_3$. In fact, the earlier as well as the present work has shown clearly that ferromagnetism emerges in nanoscale and the magnetization enhances by orders of magnitude - e.g., from $\sim$2.0 emu/cm$^3$ in bulk system to $\sim$200 emu/cm$^3$ in thin film or nanoscale systems. Magnetization enhances due to suppression of or incomplete spin cycloid, enhanced canting angle, and influence of lattice misfit strain in nanoscale systems. Theoretical calculations \cite{Wesselinowa,Jiang} show the influence of lattice misfit strain on the magnetization. With the decrease in film thickness the misfit strain enhances and this, in turn, could give rise to large rise in $M_S$ ($>$200 emu/cm$^3$) in thinner films ($<$50 nm) \cite{Jiang}. The ZFC and FC magnetization ($M$) versus temperature ($T$) data [Fig. 3(c)] were recorded across 5-300 K under 500 Oe field. Clear anomalous features (such as peaks) could be observed below $\sim$150 K in the ZFC and FC $M-T$ plots. The uncertainty in the magnetization data is around 1\%. It is important to point out here that the SrRuO$_3$ (which is used for preparing a buffer layer in between the BFO film and the STO substrate) exhibits a magnetic transition around 160 K. However, the thickness of the buffer SRO layer used here is small ($\sim$10 nm). No long-range magnetic order has been observed in such a thin SRO layer \cite{Kaur}. Therefore, influence of the magnetization of SRO layer on the magnetic properties of BFO film is not significant. We further examined whether the magnetic phase forms below 150 K is stable or not. Magnetic relaxation data were recorded over a time scale of $\sim$4000s after reaching the temperature of study ($\sim$110 K) following two different paths - (i) by bringing the sample temperature down to 110 K from 300 K under zero field and (ii) by reaching first 5 K from 300 K and then by heating the sample to 110 K under zero field. The results are shown in the supplemental materials document \cite{supplementary}. Clearly, no significant relaxation could be observed in both the cases. This observation shows that the magnetic phase is not metastable and, therefore, not dependent on the path followed in arriving below 150 K. All these results signify a genuine thermodynamic magnetic transition below 150 K. Repeated thermal cycling across 150 K too, does not have any influence on the transition features observed in ZFC and FC $M-T$ data. To the best of our knowledge, similar data for the bulk, thin film, or nanoscale BiFeO$_3$ have not been reported so far by others.

\begin{figure}[ht!]
\begin{center}
   \subfigure[]{\includegraphics[scale=0.20]{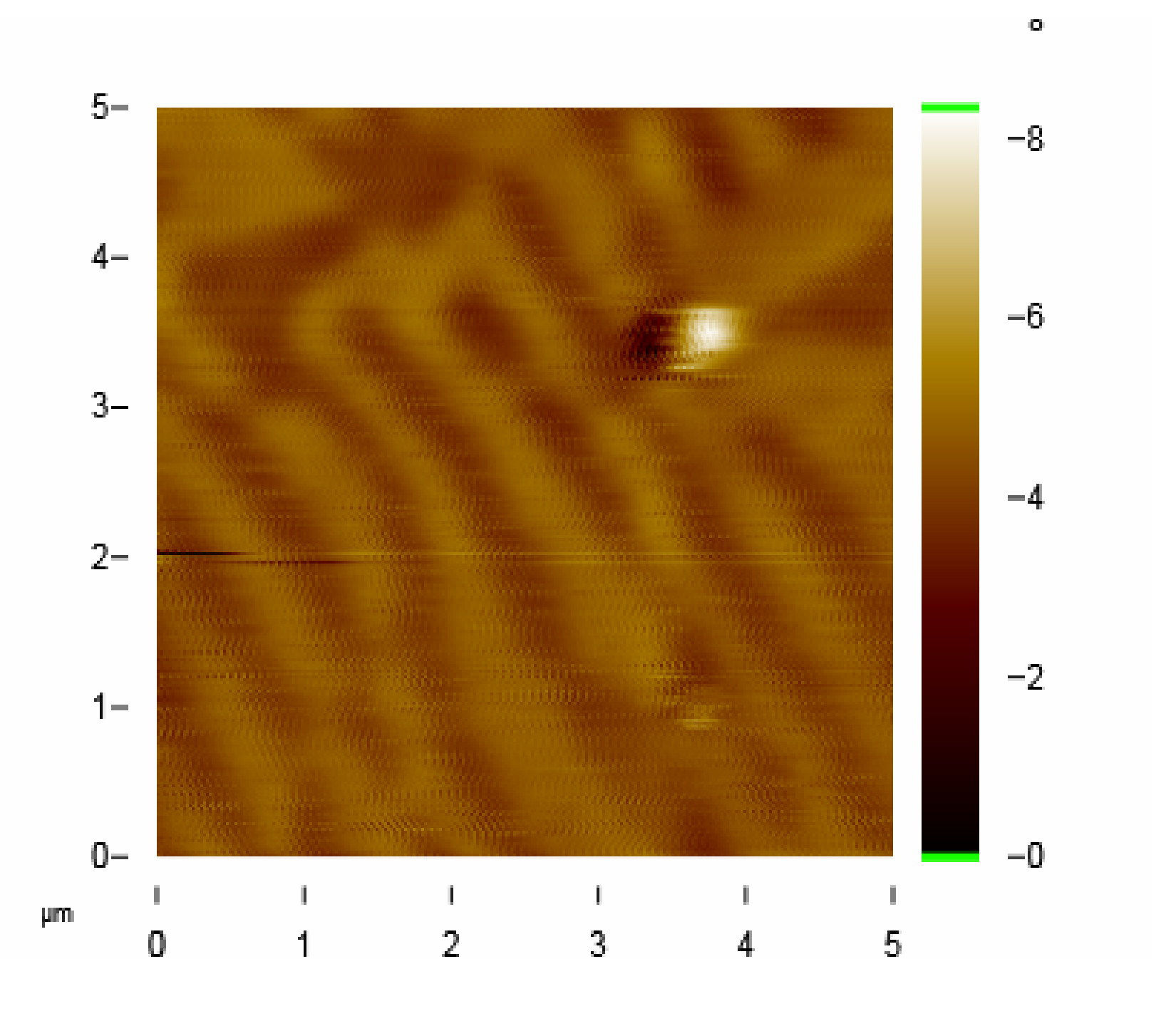}} 
   \subfigure[]{\includegraphics[scale=0.20]{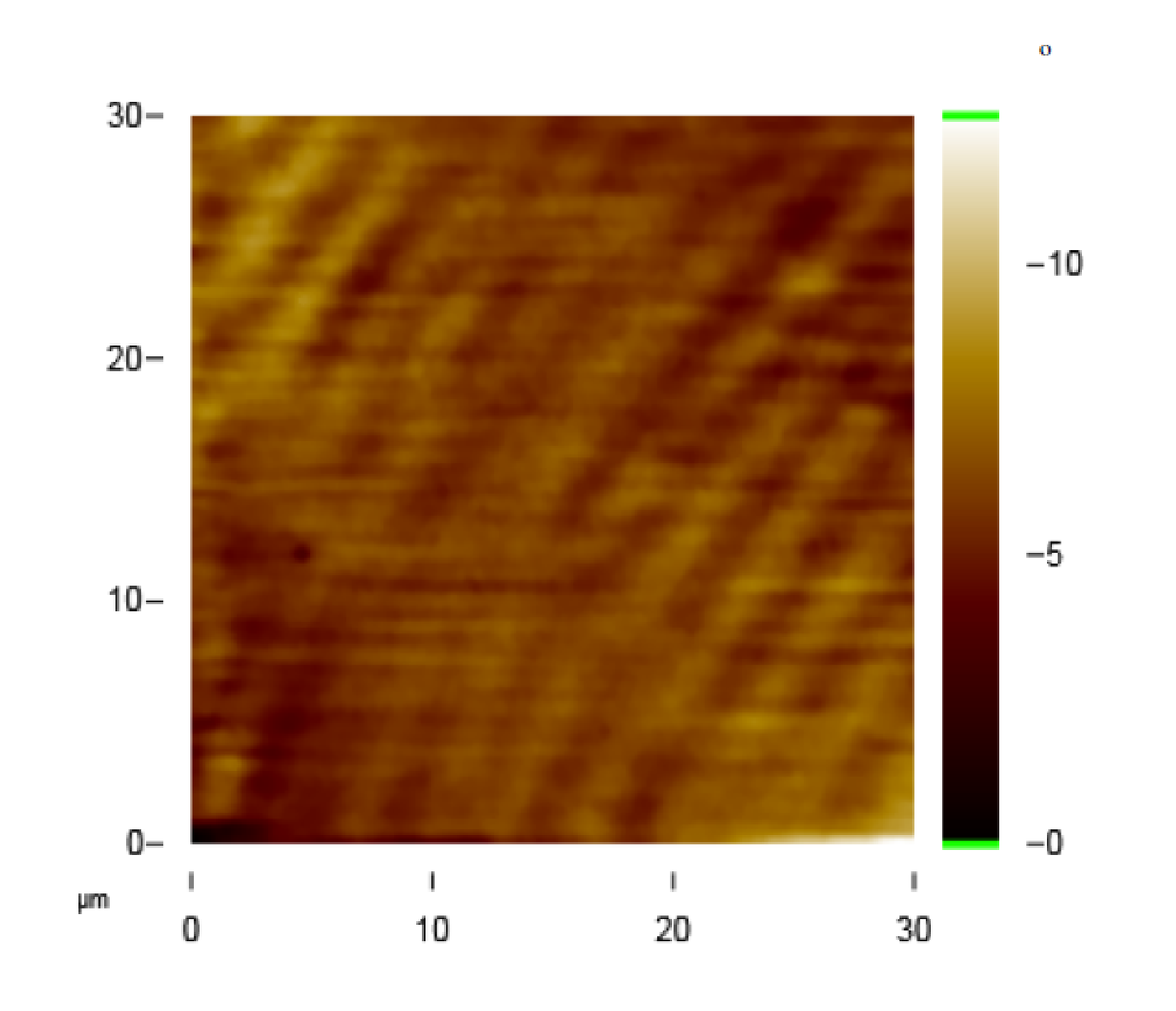}}  
   \end{center}
\caption{The phase-contrast MFM images at (a) below ($\sim$50 K) and (b) above ($\sim$200 K) the magnetic transition temperature ($T^*$ $\sim$150 K); the images have been recorded by scanning the same region of the film surface using vertical MFM (idential start coordinates of the surface for scanning have been used) under zero magnetic field; supplemental materials document \cite{supplementary} includes the line scan data taken on the images to show the domain size at above and below the transition.    }
\end{figure}

We next used magnetic force microscopy (MFM) at room temperature to image the magnetic domains and their switching characteristics under sweeping magnetic field. The commercial Co-alloy coated MFM cantilevers [Point Probe Plus Magnetic Force Microscopy - Reflex coating (PPPMFMR), nominal coercivity of approximately 300 Oe] were used for the experiments. The details of the experiments are available in an earlier paper \cite{Goswami-1}. The vertical MFM probe has been used with magnetic field applied perpendicular to the film surface, i.e., $H$ $\parallel$ [100]. It is important to point here that often the AFM and MFM images contain noises arising out of mechanical vibration and/or electrical sources. Such noises result in artifacts in the images. We have used vibration isolation bench in order to eliminate the noise due to mechanical vibration and also ensured complete elimination of the electrical noise. We repeated the measurements for ensuring observation of the intrinsic features of the magnetic domains. In Fig. 4, we show the phase-contrast MFM images. These images have been processed by WsxM software to observe the magnetic domain structure clearly. The magnetic field has been swept from zero to +70 kOe and then to -70 kOe and, finally, back to +70 kOe in order to complete the tracing of the magnetic hysteresis loop. Snapshots under different magnetic field during the tracing of forward and reverse branches of the hysteresis loop are shown in Fig. 4. The line profile data which signify 180$^o$ domain reversal upon reversal of the sign of the applied field from +70 kOe to -70 kOe are shown in Fig. 4(i). The superposition of the domain structure images [Fig. 4(h)] under +70 kOe and -70 kOe show clearly the switched domains. BiFeO$_3$ with $R3c$ space group contains the magnetic moment in the (111) plane. The vertical MFM measured along [100] exhibits the magnetic domains contained in the (100) planes, i.e., the component of the moment along [100]. The switched volume of the magnetic domains has also been estimated from the quantitative analysis of the images. Using the switched volume data, the magnetic hysteresis loop could be constructed [Fig. 4(j)]. This result shows nonlinearity in the switched domain volume versus field pattern and thereby indicates ferromagnetism. The line scans on the MFM images recorded under zero and +70 kOe field have also been recorded \cite{supplementary}. The data show increase in the domain size because of applied magnetic field \cite{supplementary}. However, there is a difference in the results of global as well as local magnetometry. While the global magnetometry exhibits saturation of magnetization within a low field regime ($\sim$15 kOe), complete saturation and thereby formation of a single domain from merging of the striped domains could not be observed in the local magnetometry carried out by MFM. This could be because of (a) complex magnetic structure in nanoscale BiFeO$_3$ comprising coexisting ferromagnetic and antiferromagnetic orders resulting from suppression of or incomplete spin spiral \cite{Bertinshaw,Gervits} and thereby difference between the global and averaged out magnetic structure and the local structure \cite{Frandsen}, (b) difference in the direction of the applied field - global magnetometry were carried out with applied field parallel to the film surface while the MFM was done with field perpendicular to the film surface. The orientation of the magnetic moment for antiferromagnetic and ferromagnetic components with respect to the crystallographic axes, however, could not be determined in the present case. It requires detailed neutron diffraction experiments. Therefore, the exact orientation of the ferromagnetic and antiferromagnetic components of the spin structure with respect to the direction of the applied magnetic field could not be ascertained. This issue will be addressed in a separate work.       

\begin{figure*}[ht!]
\begin{center}
   \subfigure[]{\includegraphics[scale=0.13]{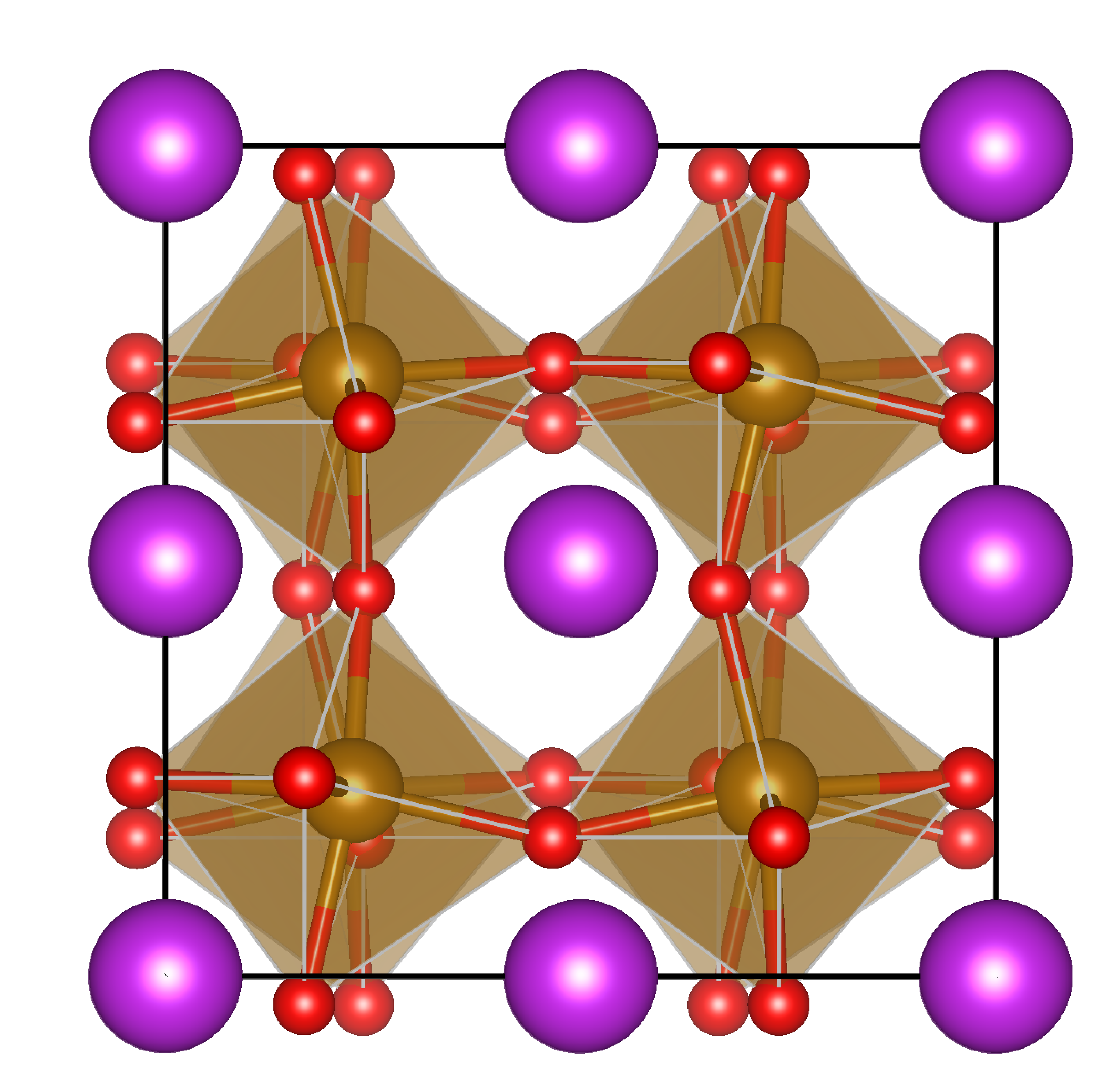}}  
   \subfigure[]{\includegraphics[scale=0.17]{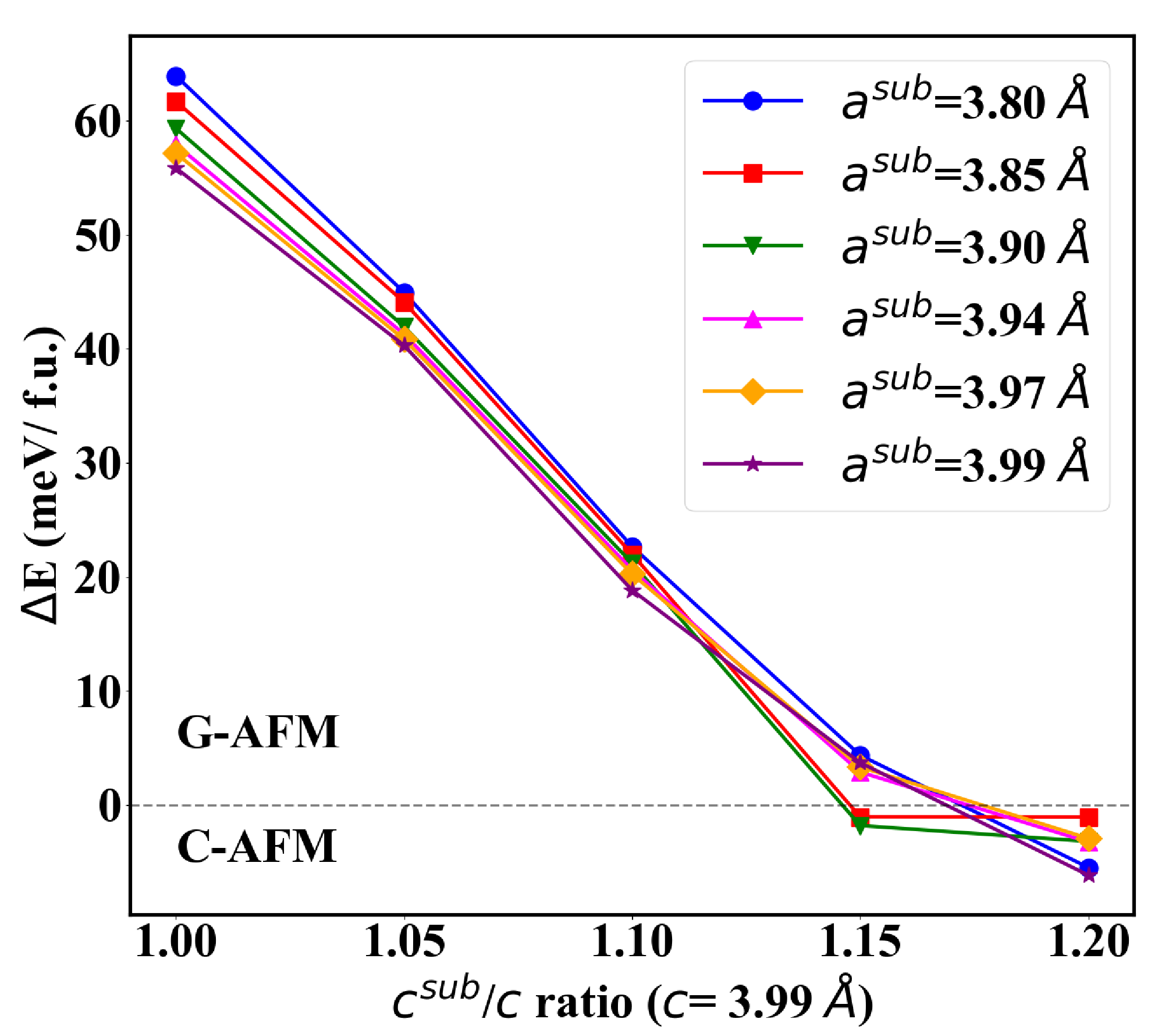}} 
   \end{center}
\caption{(a) The 2 $\times$ 2 $\times$ 2 supercell of pseudocubic BiFeO$_3$; (b) the variation of the total energy difference $\Delta E$ between the G-AFM and C-AFM structure as a function of the out-of-plane tensile strain ($c/a$ ratio) for different in-plane lattice parameter $a$.}
\end{figure*}

We further carried out the temperature-dependent MFM in order to examine whether any signaure of magnetic transition at $\sim$150 K could be detected by MFM. The film has been cooled from room temperature to the temperature of study under zero field. The phase-contrast images were then captured using vertical MFM. In Fig. 5, we show the phase-contrast MFM images recorded at $\sim$50 and $\sim$200 K. The domains are contained in the (100) plane with the moment vector aligned along [100]. It is clearly observed that, in between 200 and 50 K, the stripe-like domains have rotated by nearly 90$^o$ from [100] to [010] under zero field. This is the first evidence - obtained from MFM - of magnetic domain rotation as a result of spin-reorientation transition and consequent rotation of magnetic anisotropy (arises due to spin-orbit coupling) around 150 K in an epitaxial thin film of BiFeO$_3$. The scan area of the images differs because of the temperature-dependent piezoelectric constants of the scanner; it restricts the scan area at low temperature \cite{supplementary}. It is important to mention here that all these temperature-dependent MFM scans were recorded at the same region of the sample surface. The starting coordinates of the scanning was kept identical in all the cases. Similar images of domain rotation resulting from spin-reorientation transition under magnetic field was earlier captured \cite{Milde} by MFM for a different compound - a single crystal of Cu$_2$OSeO$_3$. We, however, did not study the influence of magnetic field on the domain rotation across 50-200 K as we focused solely on the temperature-driven spin-reorientation transition in this work. The line scans taken across the images (at $\sim$50 and $\sim$200 K) have been shown in the supplemental materials document \cite{supplementary}. The line scan data show that the spin-reorientation transition has resulted in formation of finer domains - a single domain at $\sim$200 K has been split into three at $\sim$50 K. This could possibly be because of smaller exchange coupling energy of the magnetic order emerged below 150 K. The magnetic domain size depends on the competition among several energy scales - exchange coupling, magnetostatic, magnetoelastic, magnetocrystalline anisotropy etc \cite{Cullity}. Smaller exchange energy of the spin order emerged below 150 K (transition temperature of the magnetic order prevailing above 150 K is $\sim$600 K) could give rise to smaller domain size. Similar variation in domain size as a result of spin-reorientation transition in thin films of different thickness has been noticed by others as well in Fe-Co alloy films \cite{Zdyb}. In the present case, however, as the temperature drops further (from $\sim$50 K to $\sim$3.6 K), the domain size rises again \cite{supplementary}. Earlier work \cite{Singh-1,Singh-2,Singh-3,Scott-1,Scott-2} on single crystals did observe a signature of magnetic transition near 140 K in the Raman data. The magnonic excitations at 10-30 cm$^{-1}$ were shown to exhibit anomalous slowing down and softening around 140 K. By comparing these observations with those made in the rare-earth orthoferrites it was conjectured that the anomalous features possibly originate from spin-reorientation transition. Direct measurement of the dc magnetization or ac susceptibility did not, however, offer any clear indication of spin-reorientation transition around 150 K. Instead, onset of spin glass transition could be noticed around 250 K with $T_B$ or $T_G$ around 30 K. Later work using neutron diffraction \cite{Scott-2,Ramazanoglu} also did not provide any clear evidence of reorientation of the projection of spin cycloid onto the planes studied. Even the Mossbauer spectroscopy, recorded on a thin film of BiFeO$_3$, indicated collinear antiferromagnetic structure with no spin cycloid and did not show any signature of magnetic transition in between 90 to 620 K \cite{Bibes}. It was surmised that the freezing of the magnon modes observed around 140 K is not associated with reorienation of the average projection of the spins. Using the results obtained from single crystals and nanotubes, it was pointed out \cite{Scott-3} that the anomalous Raman features as well as the change in the lattice parameters and charge density are actually surface phenomena. They are observed only in nanotubes but not in single crystals. However, powder neutron diffraction, carried out on nanoparticles of BiFeO$_3$, showed \cite{Goswami-2} emergence of different magnetic peak structure below 150 K. The magnetic structure model yields canting of the spins away from the c-axis by nearly 6$^o$.

The possibility of strain-driven magnetic transition in BiFeO$_3$ has been investigated theoretically too. The calculations were carried out on a 2 $\times$ 2 $\times$ 2 supercell [Fig. 6(a)] of pseudocubic BiFeO$_3$ (eight formula units). Bulk BiFeO$_3$ is known to assume pseudocubic rhombohedral $R3c$ structure under ambient conditions \cite{Spaldin}. We consider four different magnetic structures, namely, G-antiferromagnetic (G-AFM), C-antiferromagnetic (C-AFM), A-antiferromagnetic (A-AFM) and ferromagnetic (FM) in our calculations. First, we start with different magnetic ground states in our GGA+U calculations. G-AFM is found to be most stable with total energy lower than that for C-AFM by $\sim$55 meV/formula unit (f.u.). The FM and A-AFM structures, on the other hand, are unfavorable with higher total energy difference ($\sim$193 meV/f.u. and $\sim$120 meV/f.u., respectively). Therefore, in this work, mostly a competition between G- and C-type structures could be observed. The optimized lattice parameters $a$ = $b$ = $c$ = 3.99 $\AA$ and $\alpha$ = $\beta$ = $\gamma$ = 89.34$^o$ matches well with the previous results \cite{Dieguez,Catalan,Schmidt}. Our calculated bandgap $\sim$2.13 eV, of course, is somewhat lower than the experimental results ($\sim$2.60-2.70 eV) \cite{Ihlefeld,Basu}.

To analyze the low temperature magnetic phase transitions, we impose biaxial strains (different along the ab-plane and c-axis) and analyze mainly the two (C-AFM and G-AFM) magnetic configurations. We apply tensile strain along the c-axis of BiFeO$_3$ and compressive strain along the ab-plane, similar to our experimental results. We denote strained lattice parameters as $a^{sub}$ (modified in-plane lattice parameters $a$ = $b$) and $c^{sub}$ (modified out-of-plane lattice parameter). In essence, our new parameter $c^{sub}$ ($a^{sub}$) is larger (smaller) than $c$($a$) of the pseudocubic lattice discussed above. We calculate the energy difference between the C-AFM and G-AFM magnetic structures [$\Delta E$ = $E$(C-AFM)-$E$(G-AFM)] vs $c^{sub}$/$c$ for various $a^{sub}$ as shown in Fig. 6(b). When the lattice parameter $c^{sub}$ is left unaltered (i.e., $c^{sub}$ = 3.99 \AA), the compound assumes G-AFM ground state irrespective of the extent of compressive strain along the ab-plane. On the other hand, as the tensile strain increases along the c-axis, the $\Delta E$ gradually decreases. A crossover from G-AFM to C-AFM ground state is reached at $c^{sub}$/$c$ = 1.15 when the compressive strain is more than 2.25\% along the ab-plane. The ground state becomes C-AFM at $c^{sub}$/$c$ = 1.20. The theoretical calculations, therefore, point out that the biaxial strain in BiFeO$_3$ leads to near degeneracy of the G-AFM and C-AFM phases which paves the way for transition from G- to C-AFM structure.

\begin{figure*}[ht!]
\centering
{\includegraphics[scale=0.55]{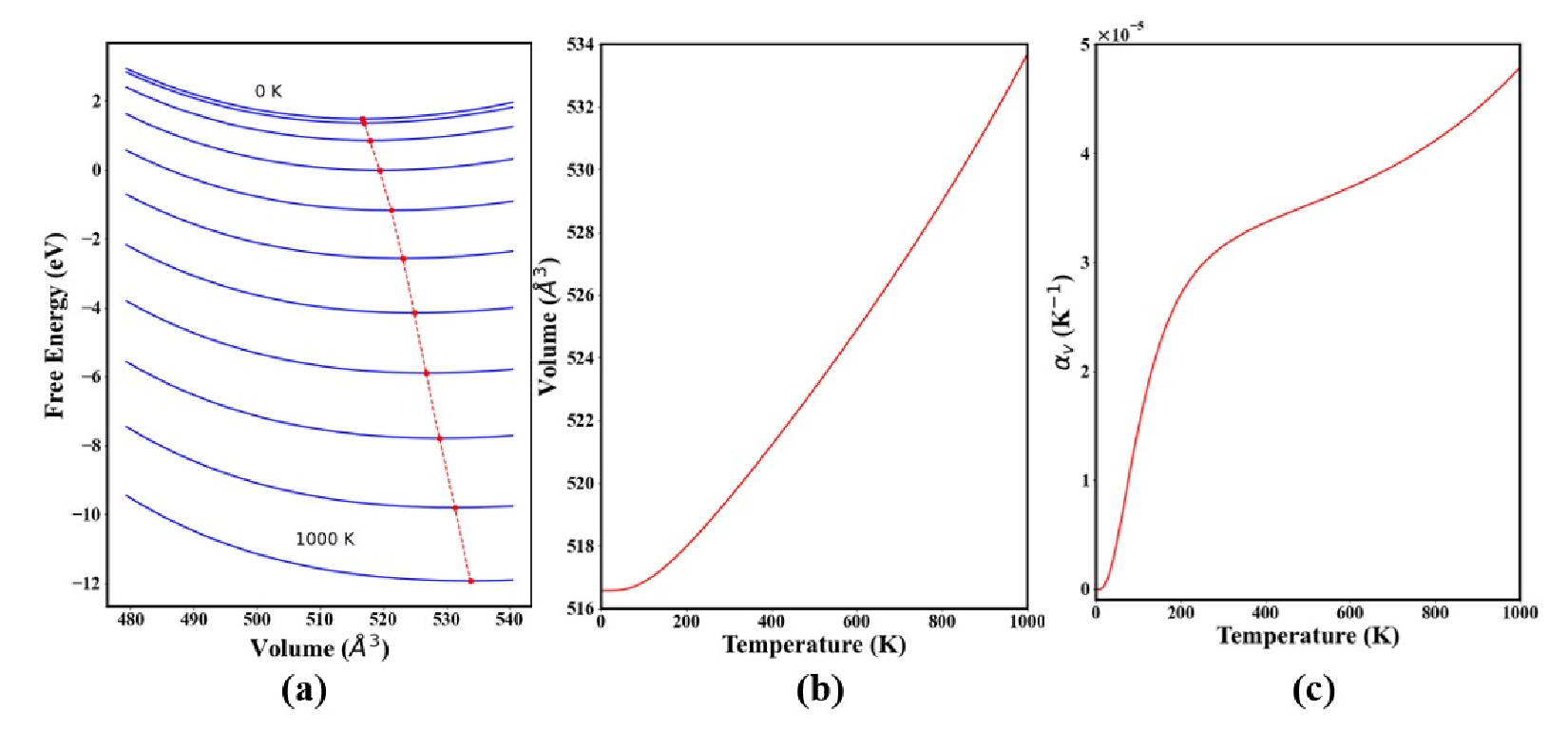}}
\caption{Quasi-harmonic approximation calculations to evaluate the equilibrium volume and thermal expansion coefficient of BiFeO$_3$ ($R3c$ phase): (a) calculated free energies with different cell volumes and temperatures; (b) equilibrium cell volume of BFO at different temperatures [obtained from red dotted line in (a)]; (c) volumetric thermal expansion coefficient ($\alpha_v$) with temperature.}
\end{figure*}

\begin{figure*}[ht!]
\centering
{\includegraphics[scale=0.55]{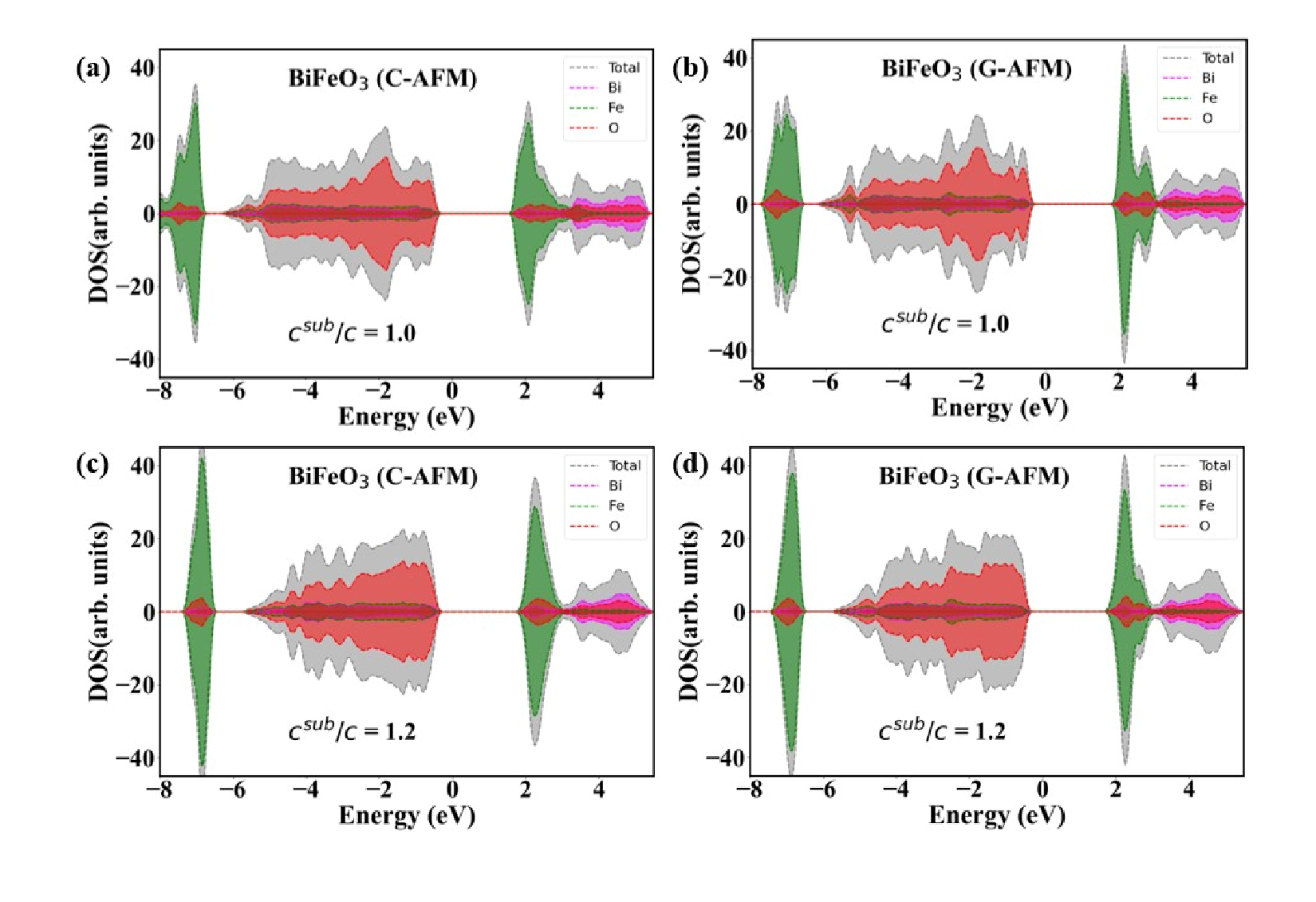}}
\caption{The density of states (DOS) calculated for the G- and C-AFM phases at unstrained and strained conditions.}
\end{figure*}

\begin{figure}[ht!]
\centering
{\includegraphics[scale=0.30]{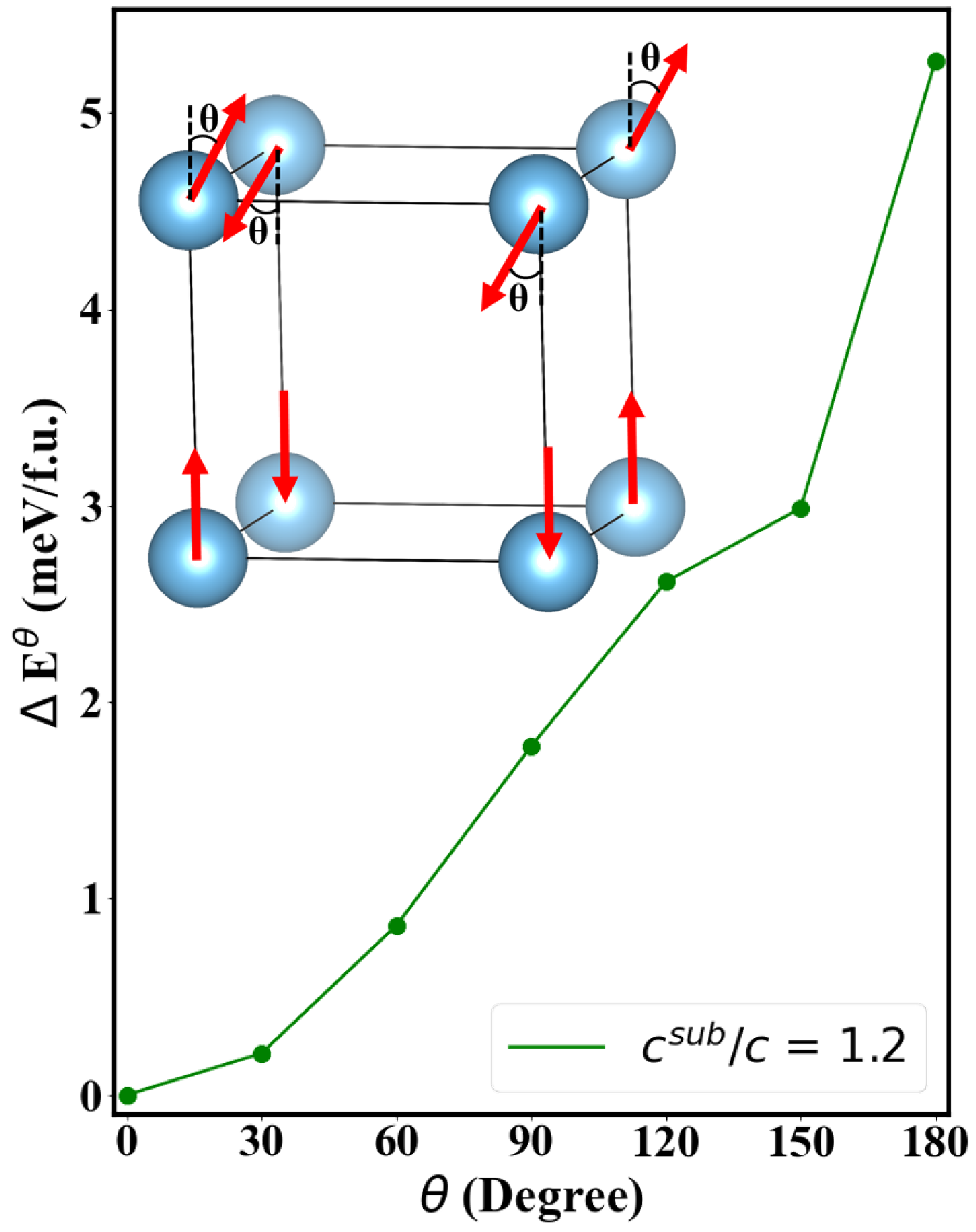}}
\caption{Energy difference $\Delta E^{\theta}$ with respect to the canting angle $\theta$ for strained BFO ($c^{sub}/c$ = 1.2 and $a$ = $b$ = 3.99 \AA) system; inset shows the schematic of the spin structure with relative canting angle $\theta$.}
\end{figure}

In order to correlate the strain-driven magnetic transition with the transition observed experimentally around 150 K, we consider the volumetric thermal expansion of the lattice. The volumetric thermal expansion coefficient is calculated using quasi-harmonic approximation as implemented in the PHONOPY package \cite{Tanaka} and the force constants of each strained supercell are calculated using density functional perturbation theory method implemented in the VASP code.

The method for obtaining the thermal expansion coefficient and subsequent lattice parameters for BFO up to a temperature $T$ = 1000 K is depicted in Fig. 7(a-c). Fig. 7(a) shows the change in the free energy with different cell volume across a temperature range 0-1000 K. Here, at a given temperature, the total free energy can be calculated using the following expression \cite{Hug}: $F(T,V) = [U_{el}(V) - U_{el}(V_0)] + F_{ph}(T,V)$ where $U_{el}(V)$ and $U_{el}(V_0)$ are the electronic ground state energies of equilibrium volume at a temperature $T$ and 0 K, respectively. Thus, the difference between these two energies represents the relative energy. $F_{ph}$ is the phonon Helmholtz free energy. 

The red dotted line [Fig. 7(b)] represents the equilibrium volume of the BFO unit cell at different temperature. The discrete calculated data is well fitted with Vinet's equation of states \cite{Vinet}. The equilibrium cell volume remains nearly constant within the temperature range 0-100 K and increases gradually between 100-200 K. A linear dependency of volume of the unit cell with temperature beyond 200 K is observed. On the other hand, volumetric thermal expansion coefficient ($\alpha_v$) varies approximately linearly in the low temperature range [Fig. 7(c)]. At higher temperature, it follows a quadratic pattern. From the above results it is clear that the equilibrium cell volume remains almost constant between 0-100 K and starts increasing within the temperature range 100-200 K. So our calculations, carried out at T = 0 K using the relaxed cell parameters, remain nearly valid till 100-150 K and, therefore, relevant for describing the magnetic transition observed experimentally around 150 K.

To further elucidate the microscopic origin of the magnetic phase transition under biaxial strain, the total and projected density of states of both the G-AFM and C-AFM phases (at $c^{sub}$/$c$ = 1.0, $a$ = $b$ = 3.99 \AA and at $c^{sub}$/$c$ = 1.20, $a$ = $b$ = 3.99 \AA) have been calculated [Figs. 8(a),(b) and 8(c),(d)]. The deep valence band region (-7.5 eV to -6 eV) is mostly dominated by the $3d$ orbitals of Fe and are relatively delocalized when no strain is applied [Fig. 8(a),(b)]. On the other hand, O-2p states mostly contribute to the top of the valence band along with minor contributions from Bi-6p states and Fe-3d states. The lower conduction band region is dominantly occupied by Fe-3d states with little contribution from O-2p or Bi-6p states. Since double exchange ferromagnetic interaction among the Fe ions is not feasible in BiFeO$_3$ \cite{Zener,Kubler,Pradhan-1,Pradhan-2} due to the absence of charge disproportionation of the transition metal ion, the superexchange interaction between the nearest neighbor Fe ions plays a vital role in determining the magnetic ground state \cite{Kanamori,Goodenough}. The strength of the antiferromagnetic superexchange interaction between the nearest neighbor Fe atoms (mediated via O atoms) is equal along all the directions leading to the G-AFM ground state. However, when the c-axis is elongated, the Fe-O-Fe bond length along the c-axis is increased, hence the overlap between Fe and O decreases. In fact, the Fe 3d orbitals in the deep valence band region (around -7 eV) are more localized as compared to the previous case [Figs. 8(c),(d)]. This localization indicates that the hopping amplitude between Fe 3d and O 2p orbitals has decreased. This, in turn, reduces the strength of the superexchange interaction between neighboring Fe ions along the c-direction. As a result, a weaker antiferromagnetic coupling between Fe atoms along out-of-plane direction as compared to the in-plane direction is expected. In such a scenario, the next nearest neighbor interaction between Fe ions starts to contest with the nearest neighbor interaction \cite{Duan}. Hence, the energy difference between G-AFM and C-AFM decreases. So, overall, a competition between the C-AFM and G-AFM configuration emerges due to the variation of the superexchange interaction along the out-of-plane direction in presence of strain and C-AFM phase turns out to be the ground state for $c^{sub}$/$c$ = 1.20.

We also investigate the influence of non-collinearity of the spin structure on these results by including the spin-orbit coupling for $c^{sub}$/$c$ = 1.20, $a$ = $b$ = 3.99 \AA. We consider variation of the spin canting angle $\theta$ in the spin structure of different planes as shown in the inset of Fig. 9. For $\theta$ = 0, our calculations reproduce the C-AFM phase to be the ground state while, for $\theta$ = 180$^o$, the ground state turns out to be G-AFM. The energy difference $\Delta E^{\theta}$ = $E(\theta)$ - $E(C-AFM)$ for different non-collinear magnetic structure has been plotted as a function of $\theta$ (Fig. 9). It is apparent that the C-AFM phase remains the ground state. But since the energy difference between the collinear and non-collinear spin structure is very small it is quite possible that because of unaccounted strain and surface effects, finite spin canting could be observed in the experimental results.       

To summarize, we point out that (i) the FM and A-AFM structure is highly unlikely to stabilize in the strain-free BiFeO$_3$, (ii) the competition is restricted between C- and G-AFM structures with rather small energy difference between the two phases ($\sim$55 meV/f.u.); (iii) introduction of biaxial strain drastically reduces the energy difference even further and clearly indicates the possibility of magnetic phase transition from G- to C-AFM structure as overlap of Fe 3d and O 2p states across the c-axis weakens in presence of strain. Small discrepancy between the theoretical and experimental results on strain-driven magnetic transition has, however, been observed. This could arise because of the possibilities of the presence of surface defects and/or spatial variation in the strain which could not be accurately accounted for by the experiments performed. It is also important to point out that to correlate the spin structure with the domain reorientation in the actual experiment, it is necessary to perform detailed neutron diffraction experiments which would provide information regarding the orientation of the spins (both ferromagnetic and antiferromagnetic components) with respect to the crystallographic structure in G- and C-states (and thereby with respect to the direction of the applied field) and the canting angles for both the states. Using these details, one can correlate the spin structure, crystallographic axes, and the 90$^o$ magnetic domain reorientation observed experimentally and also refine the model structure considered in theoretical calculations in order to make the model more realistic. This is beyond the scope of the present work and would be taken up in near future.  

The present work, therefore, offers the first convincing proof of the spin-reorientation transition in an epitaxial thin film of BiFeO$_3$ of thickness $\approx$36 nm. In order to verify whether film thickness has a role to play, we carried out ZFC and FC magnetization versus temperature measurements on a thicker film (thickness $\approx$ 120 nm). Very interestingly, the characteristic transition features \cite{supplementary} were found to have shifted to even lower temperature ($\sim$120 K). We also measured the ac susceptibility and observed the characteristic frequency-independent features around 120 K \cite{supplementary}. All these results - obtained from single crystals, nanotubes, nanoscale particles, epitaxial thin films - possibly indicate that the spin-reorientation transition results from enhanced surface to volume ratio and strain (either epitaxial in an epitaxial thin film or microstrain in a nanotube or nanoscale particle). The spin canting enhances because of the enhanced surface effect and strain which induce such a transition around 150 K. With the rise in the film thickness, influence of surface and strain weakens and hence the transition temperature shifts to low temperature regime and eventually vanishes in a single crystal. 

\section{Summary}
In summary, we show from detailed magnetometry and magnetic force microscopy that spin-reorientation transition and consequent rotation of the anisotropy results in 90$^o$ rotation of perpendicular magnetic domains around 150 K in epitaxial thin film of BiFeO$_3$ containing biaxial strain. The DFT based first-principles calculations indicate that biaxial lattice strain induces the magnetic transition from G- to C-type structure. Of course, both the G- and C-type spin structures are associated with finite spin canting and hence contain ferromagnetic and antiferromagnetic components. Since their orientation with respect to the crystallographic axes and hence applied magnetic field could not be ascertained, direct correlation between the observed 90$^o$ domain rotation and the theoretically calculated transition from G- to C-type spin structure below 150 K could not be established. The strain-driven effect, however, weakens with the increase in film thickness possibly because of weaker influence of strain and, hence, a `strain-free' single crystal does not exhibit the transition.  

\begin{center}
$\textbf{ACKNOWLEDGMENTS}$
\end{center}

This work is supported by the project SPR/2021/000131 sponsored by the Science and Engineering Research Board (SERB), Government of India. One of the authors (D.M.) expresses gratitude for the financial support received from the Technical Research Center (TRC), Department of Science and Technology (DST), Government of India (Grant No. AI/1/62/IACS/2015). Another author (S.C.) acknowledges fellowship from DST-INSPIRE, Government of India.

\end{document}